\documentclass[superscriptaddress,
amsmath,amssymb,
aps,
pra,
floatfix,
twocolumn]{revtex4-2}

\usepackage[utf8]{inputenc}
\usepackage[english]{babel}
\usepackage[T1]{fontenc}
\usepackage{hyperref}
\usepackage{amsmath}
\usepackage{amssymb}
\hypersetup{colorlinks=true,
    linkcolor=blue,
    filecolor=blue, 
    citecolor=blue,
    urlcolor=blue,
}

\usepackage{braket, hyperref, mathtools, verbatim}
\usepackage{physics}
\usepackage{graphicx}
\usepackage{braket}
\usepackage{dcolumn}
\usepackage{bm}
\usepackage{tabularx}
\usepackage{makecell}
\usepackage{tikz}
\usepackage{lipsum}
\usepackage{amsmath}
\usepackage{amssymb}
\usepackage{bbold}
\usepackage{algorithm2e}
\usepackage{enumerate}
\usepackage{mathtools}
\SetKwComment{Comment}{/* }{ */}
\RestyleAlgo{ruled}
\SetAlgoCaptionLayout{raggedright}

\newcommand{\Z}{\mathbb{Z}}

\newcommand{\K}{\mathbb{K}}

\newcommand{\sL}{\mathcal{L}}

\newcommand{\V}{\mathcal{V}}
\newcommand{\sD}[0]{\mathcal{D}}

\newcommand{\had}{\hat{a}^{\dagger}}
\newcommand{\ha}{\hat{a}}
\newcommand{\hhd}{\hat{h}^{\dagger}}
\newcommand{\hh}{\hat{h}}

\newcommand{\td}[1]{\tilde{#1}} 
\newcommand{\nbar}[0]{\bar{n}} 
\newcommand{\mbf}[1]{\mathbf{#1}} 
\newcommand{\hSig}[0]{\hat{\sigma}} 
\newcommand{\hTau}[0]{\hat{\tau}}

\newcommand{\cfb}[0]{\overline{C}_4}
\newcommand{\exb}[0]{\mathbf{E}\times\mathbf{B}}
\newcommand{\hHo}{\hat{H}_\text{odf}}
\newcommand{\hUo}{\hat{U}_\text{odf}}

\newcommand{\hU}{\hat{U}}
\newcommand{\hH}{\hat{H}}

\DeclarePairedDelimiter{\dKet}{\lvert}{\rangle\!\rangle}
\DeclarePairedDelimiter{\dBra}{\langle\!\langle}{\rvert}
\newcommand{\dmel}[3]{\langle\!\langle #1 \lvert #2 \rvert #3 \rangle\!\rangle}
\newcommand{\dip}[2]{\langle\!\langle #1 \vert #2 \rangle\!\rangle}
\newcommand{\dop}[2]{\lvert #1 \rangle\!\rangle \langle\!\langle #2 \rvert}
\newcommand{\dev}[2]{\langle\!\langle #2 \lvert #1 \rvert #2 \rangle\!\rangle}

\begin{document}

\title{Bilayer crystals of trapped ions for quantum information processing}

\begin{abstract}
 Trapped ion systems are a leading platform for quantum information processing, but they are currently limited to 1D and 2D arrays, which imposes restrictions on both their scalability and their range of applications. Here, we propose a path to overcome this limitation by demonstrating that Penning traps can be used to realize remarkably clean bilayer crystals, wherein hundreds of ions self-organize into two well-defined layers. These bilayer crystals are made possible by the inclusion of an anharmonic trapping potential, which is readily implementable with current technology. We study the normal modes of this system and discover salient differences compared to the modes of single-plane crystals. The bilayer geometry and the unique properties of the normal modes open new opportunities, in particular in quantum sensing and quantum simulation, that are not straightforward in single-plane crystals. Furthermore, we illustrate that it may be possible to extend the ideas presented here to realize multilayer crystals with more than two layers. Our work increases the dimensionality of trapped ion systems by efficiently utilizing all three spatial dimensions and lays the foundation for a new generation of quantum information processing experiments with multilayer 3D crystals of trapped ions.  
\end{abstract}

\newcommand{\iisc}{\affiliation{Department of Instrumentation and Applied Physics, Indian Institute of Science, Bangalore 560012, India.}}
\newcommand{\iitb}{\affiliation{Department of Physics, Indian Institute of Technology-Bombay, Powai, Mumbai 400076, India.}}
\newcommand{\jila}{\affiliation{JILA, National Institute of Standards and Technology,and Department of Physics, University of Colorado, Boulder, CO 80309}}
\newcommand{\ctqm}{\affiliation{Center for Theory of Quantum Matter, University of Colorado, Boulder, CO 80309}}
\newcommand{\nist}{\affiliation{National Institute of Standards and Technology, Boulder, CO 80309}}

\author{Samarth Hawaldar}
\email{samarthh@iisc.ac.in}
\iisc
\author{Prakriti Shahi}
\iitb
\author{Allison L. Carter}
\nist
\author{Ana Maria Rey}
\jila\ctqm
\author{John J. Bollinger}
\nist
\author{Athreya Shankar}
\email{athreyas@iisc.ac.in}
\iisc

\maketitle

\section{Introduction}

Quantum hardware is now coming full circle: Early experiments studied quantum phenomena by trapping and working with large ensembles of particles--- typically several thousands to even millions---  but with a relatively low level of control. Subsequent developments in laser cooling and trapping, as well as in solid state technology, paved the way for studies of systems with just a single or a few quantum units, and with a very high degree of control over the small numbers of degrees of freedom of these systems. In the last decade, there has been rapid progress in combining the best of the previous two eras in order to once again scale up quantum hardware, but at the same time, also  attempt to maintain a high level of controllability. Paradigmatic systems in this so-called Noisy Intermediate Scale Quantum (NISQ) era~\cite{preskill2018Quantum} are composed of several tens to hundreds of qubits arranged in 1D or 2D arrays with varied and even reconfigurable connectivity graphs, capable of reasonably high-fidelity quantum operations and measurements, and are designed to explore genuine quantum many-body features that are not easily tractable with classical computers.  Such systems have now been realized on a wide variety of platforms, including but not restricted to trapped ions~\cite{monroe2021RMP}, neutral atoms in tweezers and optical lattices~\cite{ebadi2021Nat}, quantum dots~\cite{borsoi2023NatNano}, and superconducting quantum circuits~\cite{kim2023Nat}, to name but a few.  

A further scaling up of controllable quantum hardware will benefit from an efficient use of all three spatial dimensions, e.g., by devising strategies to create multilayer structures by stacking two or more 2D arrays of particles. In addition, multilayer arrays open up new avenues for quantum sensing applications and their three-dimensional order enables them to make contact with quantum simulations of models that are relevant for condensed matter systems, including but not limited to ladder arrays, bilayer graphene, and Moir\'e materials~\cite{Senthil2015,Novoselov2004,Wang2007,Andrewi2021,Lutchyn2010,Hague2012,Tutela2019,Guijarro2022}. Several platforms have taken rapid strides in this direction, with the demonstration of synthetic 3D arrays of atoms in optical tweezers and lattices~\cite{campbell2017Sci,barredo2018Nat,kumar2018Nat,schlosser2023PRL}, layer-resolved arrays of optically trapped polar molecules~\cite{tobias2022Sci} and  fermionic atoms~\cite{Gall2021},
and, very recently, atomic Bose–Einstein  condensates~\cite{Meng2023}  in twisted-bilayer optical lattices. On the other hand, a number of other leading quantum information processing platforms appear to be limited to 1D or 2D arrays in the near term, because it remains a challenge to identify a scalable route to produce multilayer structures in these platforms.

In particular, this challenge is very evident in the case of trapped ion quantum systems, which have emerged as one of the major platforms for quantum information processing in the NISQ era~\cite{monroe2021RMP}. Typically, in these systems, a spin-$1/2$ system is encoded in two long-lived electronic states of each ion, and lasers are used to couple these electronic states to the collective normal modes of motion of the ions, which mediate spin-spin interactions. State-of-the-art quantum simulation and quantum sensing experiments have been proposed and realized with linear chains of several tens of ions in rf Paul traps~\cite{zhang2017Nat,zhang2017Nat2,kokail2019Nat,marciniak2022Nat,joshi2022Sci}, and also with planar 2D crystals of tens to hundreds of ions in both rf Paul traps~\cite{syzmanski2012APL,donofrio2021PRL,qiao2022arXiv,kiesenhofer2023PRXQ} as well as in Penning traps~\cite{britton2012nat,bohnet2016Science,garttner2017NatPhys,mcmahon2020PRA,ball2019RSI,jain2020PRX,shankar2022PRXQ}. However, studies exploring the possibility to realize and use 3D arrays of ions for quantum information have been rather sparse~\cite{schmied2009PRL,ravi2010PRA,palani2023PRA}, and moreover, their practical feasibility and/or scalability remain unclear.  

\begin{figure}[h]
	\centering
    \includegraphics[width=0.85\linewidth]{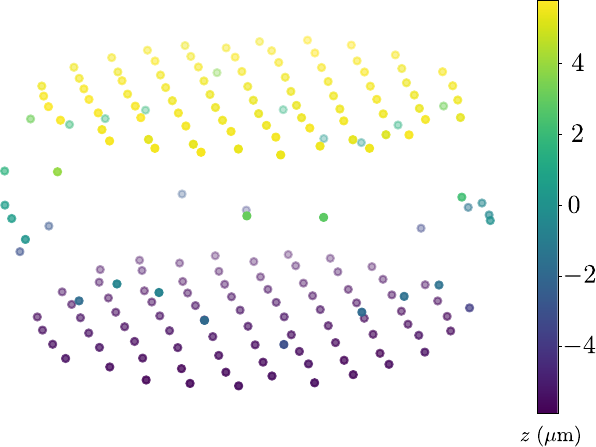}
	\caption{\textbf{Bilayer trapped ion crystals for quantum information processing.} 3D view of a numerically obtained equilibrium configuration of a bilayer crystal with $N=200$ trapped ions in a Penning trap. Markers represent ion positions, color coded according to their $z$ coordinate. The trap center (not shown) is taken as the origin. In the lab frame, the crystal is rotating with a controllable rotation frequency about a central axis that is normal to the layers, i.e., along the $z$ direction. The crystal configuration is shown in the rotating frame. The trapping conditions under which this configuration is obtained is discussed in Sec.~\ref{sec:anharmonic_eq}.}
    \label{fig:bilayer_3d}
\end{figure}

In this work, we take a step towards addressing this challenge by demonstrating that Penning traps can be used to prepare bilayer crystal configurations of hundreds of ions with well-defined layers. An advantage of Penning traps is their well-established ability to produce and control large ion crystals, and the absence of micromotion which arises in rf Paul traps because of the time-dependent trapping potentials. An example of a bilayer crystal we obtain numerically under optimal trapping conditions is shown in Fig.~\ref{fig:bilayer_3d}. 
Remarkably, such configurations do not require intricate engineering of trap structures, but are instead obtained by including an anharmonic term in the electrostatic trapping potential, which is straightforward to achieve in practice.
As we will demonstrate, the striking crystal geometry is not just a cosmetic feature; instead, the bilayer structure and the normal modes hosted by such a crystal enable opportunities for quantum information processing that are typically beyond the capabilities of 1D or 2D ion crystals. A crucial capability that the bilayer structure adds to the system is the ability to generate and detect bipartite entanglement between two spatially separated ensembles, a capability which other platforms go to great extents to achieve~\cite{fadel2018Sci,kunkel2018Sci,lange2018Sci}. Bipartite entanglement between spatially separated ensembles has been identified as a crucial resource for several quantum information processing applications~\cite{Vitagliano2023Quantum}, and furthermore, enables the demonstration of conflicts with classical physics in a very pronounced manner~\cite{fadel2018Sci}. Moreover, each layer of the bilayer system consists of a large number of spins, and hence can be effectively described using continuous quantum variables~\cite{cerf2007quantum}. This opens a new platform for continuous variable quantum simulation and, potentially, even quantum computation. Although our focus in this work is primarily on the preparation and applications of bilayer crystals, the techniques presented here may enable the preparation of crystals with more than two layers, as we briefly illustrate towards the end of our study. We note that, although 3D crystals have been realized in both rf Paul traps and Penning traps with purely harmonic trapping potentials, their geometry is strongly affected by surface effects for moderate ion numbers and only a small fraction of ions form a 3D periodic structure (see Sec.~\ref{sec:harmonic_eq}), which becomes evident only for very large ion numbers~\cite{dubin1989PRA,hasse1991PRA,tan1995PRL}. In contrast, our work shows a path for preparing crystals with several hundreds to a few thousand ions where a majority of the ions form a layered, ordered structure in 3D, thus enabling a range of quantum capabilities and applications which benefit from such a structure. 

\subsection{Scope of Work and Summary}
We undertake an extensive study of the classical and quantum physics underlying the realization of bilayer trapped ion crystals and their use in quantum information processing. In the following, we describe the different aspects encompassed by our study and highlight our major contributions.
\begin{enumerate}[(i)]
    \item We explore the possibility of using a Penning trap to realize crystals with two well-defined layers of ions, which we refer to as a bilayer crystal (Sec.~\ref{sec:equilibrium}). Although theory predicts the existence of a bilayer phase in a purely harmonic trap~\cite{dubin1993PRL,mitchell1998Science,mitchell1999POP}, we find numerically that the layers are not well demarcated due to strong boundary effects arising from a finite number of ions. Remarkably, we discover that these boundary effects can be mitigated by the addition of a quartic (anharmonic) trapping potential of optimal strength, which results in well-defined bilayer crystals even with moderate numbers of ions $\sim \mathcal{O}(100)$.
    \item Quantum information protocols with trapped ions utilize the shared motion of ions to mediate inter-ion interactions. Therefore, we characterize and quantize the normal modes of bilayer trapped ion crystals, and compare their properties with the modes of single-plane crystals (Sec.~\ref{sec:normal_modes}). A prominent difference is that the bilayer structure causes modes \emph{along} the magnetic field (axial modes) to acquire a chiral character, which is only associated with the modes \emph{perpendicular} to the magnetic field in single-plane crystals. This feature is exciting because quantum information protocols typically utilize the axial modes, whereas coupling to the perpendicular modes is challenging because of the crystal rotation. We note that chiral modes do not arise in crystals formed in rf Paul traps.
    \item The spin-motion coupling required to mediate inter-ion interactions is typically implemented by creating an optical lattice using a pair of lasers. This lattice induces a spin-dependent force on the ions, which we refer to as the optical dipole force (ODF) and study in the context of the bilayer crystal in Sec.~\ref{sec:quantum_control}. We show how the ODF provides an operationally meaningful way to assess the bilayer quality. We introduce the notion of an interlayer ODF phase, which can be tuned by changing the incidence angle of the lasers, and serves as a powerful knob for controlling the relative sign and magnitude of interlayer to intralayer ion-ion interactions.    
    \item \label{item:prospects} We discuss several prospects for quantum information processing with bilayer crystals in Sec.~\ref{sec:prospects}. We show how to realize bilayer Ising models, where the interlayer interactions can be tuned from ferromagnetic to antiferromagnetic, and even switched off, by simply adjusting the ODF incidence angle. We go on to further demonstrate that a two-tone ODF protocol, where two axial modes are simultaneously used for the spin-motion coupling, can enable \emph{on-the-fly} control over the interlayer as well as intralayer couplings. Furthermore, we consider the addition of a transverse field, which converts bilayer Ising models into bilayer spin-exchange models. Here, we demonstrate that the spin-exchange coefficients between a pair of ions $j,k$ can be made asymmetric, i.e. $J_{jk}\neq J_{kj}$, which enables the realization of chiral spin-exchange models. For example, this can be used to engineer a complex amplitude for the hopping of spin excitations between layers, opening a path for engineering a 2D synthetic gauge field \cite{Goldman2014, Dallibard2011,Zhai2015,Aidelsburger2018, Rey2016, Galitski2019} and for studying the interplay between the nontrivial topology of a band structure and inter-particle interactions.
    We also outline several other potential applications of the above capabilities in various quantum sensing and quantum simulation protocols.
    \item We identify and discuss practical challenges that can limit the fidelity of quantum protocols with bilayer crystals, and outline some strategies to mitigate their adverse affects (Sec.~\ref{sec:practical}). In particular, we find that the main challenges are likely to be off-resonant light scattering (spontaneous emission) from the ODF lasers and residual thermal motion of the ions. Our analysis suggests that the fidelity of quantum protocols using bilayer crystals may critically rely on near ground-state cooling of all the normal modes, for which we briefly discuss some prospects.  
    \item We close with a Conclusion and Outlook (Sec.~\ref{sec:conclusion}), where we demonstrate the possibility to produce multilayer ion crystals beyond bilayers. We also highlight interesting possibilities of employing bilayer crystals for simulating spin-boson models and additional simulation possibilities afforded by employing a Molmer-Sorensen gate. Finally, we underscore the need for complementary efforts in non-neutral plasma physics in the general context of utilizing 3D ion crystals for quantum information processing. 
\end{enumerate}

\section{Equilibrium Bilayer Crystals}
\label{sec:equilibrium}

In this section, we discuss the formation of bilayer crystals in a Penning trap. We first consider the case of a purely harmonic trap, introduce the system Lagrangian and briefly discuss the one-to-two plane transition. Subsequently, we study the equilibrium crystal configuration in the bilayer regime. We then demonstrate how the addition of an anharmonic trapping potential greatly improves the bilayer structure of these crystals.

\subsection{Harmonic trapping potential}
\label{sec:harmonic_eq}
\subsubsection{Preliminaries: Lagrangian of the system}
\begin{figure*}
	\centering
    \includegraphics[width=0.95\linewidth]{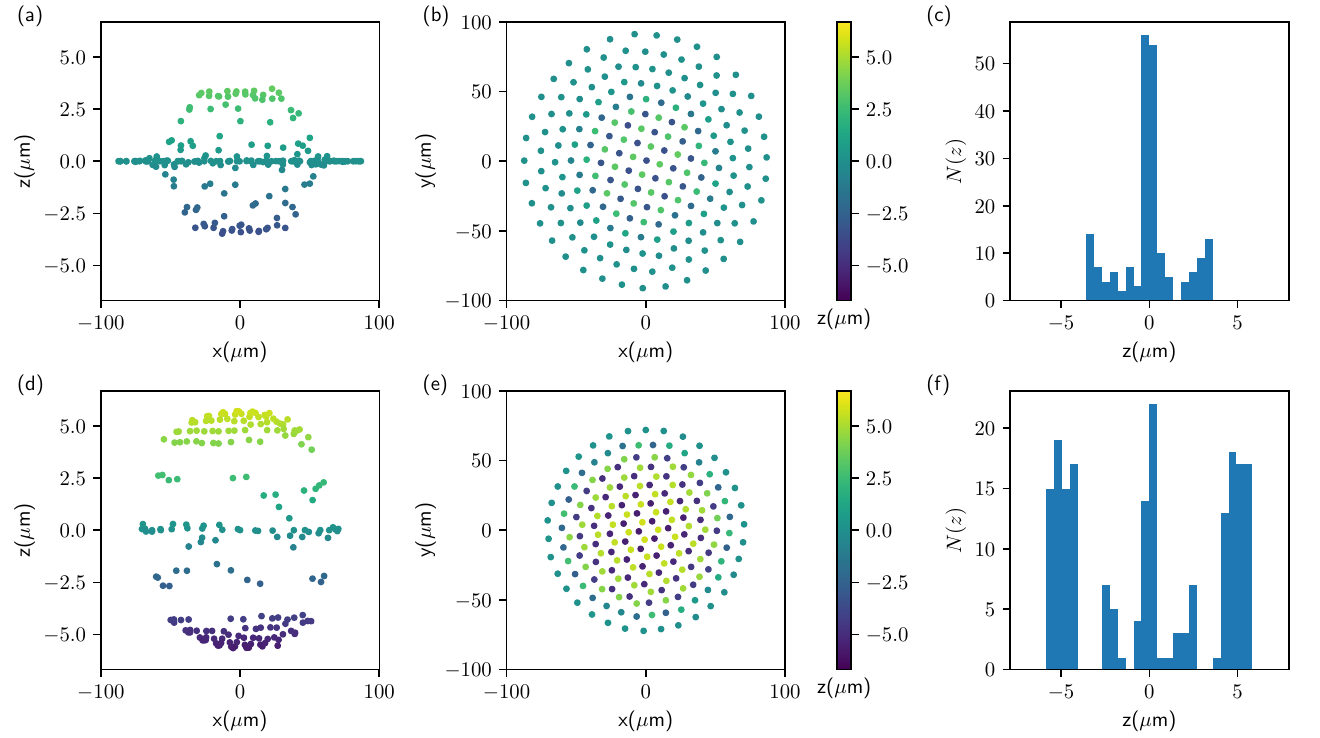}
\caption{\textbf{Bilayer crystals in a harmonic trapping potential.} Panels (a)-(c) respectively show the side view, top view and histogram of $z$ positions for a bilayer crystal with a staggered square lattice (which is evident in the central region of the top view). Panels (d)-(f) show the corresponding plots for a bilayer crystal with a staggered hexagonal lattice, which is obtained at higher rotation frequencies. Although both crystals exhibit a semblance of two layers, one each above and below the $z=0$ plane, they are not clean planar structures and are not well-demarcated, gradually merging into a single layer at $z=0$ at the crystal boundary, where a large number of ions are trapped. These crystals are obtained using parameters relevant for the NIST Penning trap [see Eqs.~(\ref{eqn:lag_full})-(\ref{eqn:delta})]: $N=200$ ${}^9\mathrm{Be}^+$ ions are trapped using an axial magnetic field of $B_z=4.4588$ T and an electic quadrupole potential characterized by the axial trapping frequency $\omega_z/(2\pi)=1.62$ MHz. For the crystal in the top row, the crystal rotation frequency is $\omega_r/(2\pi)=200$ kHz, and in the bottom row, $\omega_r/(2\pi)= 220$ kHz. The wall strength is taken to be $\delta=0.00183$.
}
    \label{fig:harmonic_eq}
\end{figure*}

We consider a collection of $N$ ions of mass $m$ in a Penning trap~\cite{wang2013PRA,shankar2020PRA}. The ions are trapped by a combination of a strong magnetic field $\vb{B}=B_z \vu{z}$ along the $z$-axis ($B_z>0$) and an electric quadrupole field characterized by a voltage amplitude $\V_z>0$ (units of $\text{V}/\text{m}^2$). In the lab frame, the combination of trapping fields causes the crystal to rotate about the $z$-axis, and the corresponding rotation frequency $\omega_r$ can be stabilized and controlled by applying a time-varying `rotating wall' potential with amplitude $\V_W>0$~\cite{huang1998PRL}. In a frame rotating at $\omega_r$ with the rotating wall, the Lagrangian is time independent and given by
\begin{align}
\label{eqn:lag_full}
\sL = \sum_{j = 1}^N \left[\frac{1}{2}m \dot{\vb{r}}_j\cdot\dot{\vb{r}}_j - \frac{m \omega_c'}{2}(\dot{x}_j y_j - \dot{y}_j x_j) - e\phi_j\right],
\end{align}
where $\vb{r}_j=x_j\vu{x}+y_j\vu{y}+z_j\vu{z}$ is the position of ion $j$ in the rotating frame, with the trap center taken as the origin. The first term describes the kinetic energy of the ions and the second term is the net Lorentz force in the rotating frame, which is characterized by the effective cyclotron frequency $\omega_c' = \omega_c-2\omega_r$, where $\omega_c=eB_z/m$ is the bare cyclotron frequency. The third term is the effective potential energy of ion $j$, which contains terms arising from the trapping potential and the inter-ion Coulomb repulsion, and is given by  
\begin{align}
\label{eqn:harmonic_phi}
e\phi_j = &\frac{1}{2}m\omega_z^2 \left[ z_j^2 + \beta (x_j^2 + y_j^2) + \delta (x_j^2 - y_j^2) \right]\nonumber\\
&+ \frac{k_e e^2}{2} \sum_{k\ne j}\frac{1}{r_{jk}},  
\end{align}
where $k_e=1/(4\pi\epsilon_0)$ with $\epsilon_0$ the vacuum permittivity. Furthermore, we have introduced the axial trapping frequency $\omega_z=\sqrt{2e\V_z/m}$ and expressed the radial trapping terms using dimensionless parameters normalized to $\omega_z^2$. In particular, the relative strength of radial to axial trapping is characterized by the parameter $\beta$, defined as~\cite{Dubin2013PRA} 
\begin{align}
    \beta = \frac{\omega_\perp^2}{\omega_z^2} = \frac{\omega_r(\omega_c-\omega_r)}{\omega_z^2}-\frac{1}{2},
    \label{eqn:beta}
\end{align}
where $\omega_\perp=\sqrt{\omega_r(\omega_c-\omega_r)-\omega_z^2/2}$ is the radial trapping frequency. The expression for $\omega_\perp$ can be understood as coming from an effective planar potential in the rotating frame that includes contributions from the Lorentz force ($\propto\omega_r\omega_c$), the centrifugal force ($\propto\omega_r^2$), and the radially outward electric quadrupole field ($\propto\omega_z^2$). Furthermore, the strength of the applied rotating wall is parameterized by the ratio $\delta$, given by~\cite{tang2021PRA} 
\begin{align}
    \delta = \frac{\V_W}{\V_z}.
    \label{eqn:delta}
\end{align}

We numerically obtain equilibrium crystal configurations by minimizing the total potential energy appearing in Eq.~(\ref{eqn:lag_full}). Starting from a random configuration of ions, we use a modified version of a basin-hopping algorithm that involves several iterations of gradient-descent based minimization interspersed with random perturbations to the ion positions to nudge the crystal out of local minima. The details of this procedure are described in Appendix~\ref{app:Modified_Basin_hopping}.

\subsubsection{One-to-two plane transition}

A single plane 2D crystal is formed when the radial trapping is sufficiently weak compared to the axial trapping. This occurs when~\cite{Dubin2013PRA,tang2021PRA} 
\begin{align}
    \beta < \beta_c \approx 0.665/\sqrt{N}.
\end{align}
For $\beta>\beta_c$, the crystal is no longer confined to a single plane. Experimentally, the transition from a single plane to a 3D crystal has been observed to occur via a series of transitions in the crystal structure, which bifurcates into an increasing number of layers as $\beta$ is increased, e.g., by increasing the rotation frequency $\omega_r$~\cite{mitchell1998Science,mitchell1999POP}. Here, we are primarily interested in the regime where the crystal structure exhibits two prominent layers.
\subsubsection{Crystal configurations}
To study a concrete realization, we consider parameters relevant for the NIST Penning trap, where single-plane crystals of tens to hundreds of ${}^9\mathrm{Be}^+$ ions are routinely prepared~\cite{bohnet2016Science}. Here, we consider $N=200$ ions trapped using a magnetic field $B_z=4.4588\,$T [cyclotron frequency $\omega_c/(2\pi)\approx 7.5973\,$MHz] and an axial trapping frequency of $\omega_z/(2\pi)=1.62\,$MHz. The critical value for the one-to-two plane transition is $\beta_c\approx 0.047$. Typically, planar crystals are prepared for quantum simulation and sensing using a rotation frequency $\omega_r/(2\pi)=180\,$kHz and wall strength $\delta\approx 1.83\times 10^{-3}$, for which $\beta/\beta_c\approx 0.186$. In the Supplemental Material, we demonstrate the one-to-two plane transition by means of an animation that shows how the crystal configuration changes as $\omega_r$ (and thus $\beta/\beta_c$) is increased.    

Here, we study representative bilayer crystals in the regime where $\beta/\beta_c>1$.  Figures~\ref{fig:harmonic_eq}(a) and~(b) show the side and top views of a crystal when $\beta/\beta_c = 1.355$ [$\omega_r/(2\pi)=200\,$kHz]. The ion positions are color coded according to their $z$ positions. The ions in the center of the crystal approximately organize into two layers, one each above and below the $z=0$ plane. With increasing distance from the trap center, the ion positions gradually shift towards the $z=0$ plane and eventually form concentric rings in this plane at the crystal boundary. As seen in the top view, the crystal has a square  lattice structure in the center. Furthermore, the lattice in the two layers are staggered with respect to each other, i.e. ions in the bottom layer are visible in the `void' created by the top layer. As a qualitative indicator of the bilayer quality, we histogram the $z$ positions of the ions in Fig.~\ref{fig:harmonic_eq}(c). The bin size is chosen keeping laser addressing in mind and will be discussed in Sec.~\ref{sec:ODF}. It is clear that the majority of the ions are still located near the $z=0$ plane, although the bilayer structure shows up as two smaller peaks close to $z=\pm 3 \; \mu$m.  

In Fig.~\ref{fig:harmonic_eq}(d-f), the system is pushed deeper into the bilayer regime by speeding up the crystal rotation to $\omega_r/(2\pi)=220\,$kHz, so that $\beta/\beta_c=2.518$. From the side view and the histogram, we observe that the stronger radial confinement pushes the ions farther away from the $z=0$ plane and reduces the number of ions trapped in this plane. The crystal now has a hexagonal lattice structure in the crystal center. (see top view). Once again, the lattice in the two layers are staggered since the ions in the bottom layer are visible in the tetrahedral `voids' formed by the top layer.

We note that the observed square lattice and the subsequent transition to the hexagonal lattice are consistent with prior experimental observations of structural phase transitions in large crystals stored in Penning traps~\cite{mitchell1998Science,mitchell1999POP}. Although theory predicts the existence of a clean bilayer phase~\cite{dubin1993PRL} for a system that is infinite in the radial direction, our simulations indicate that in a purely harmonic potential, finite size effects ($N=200$ here) cause the crystal to eventually taper into a single-plane configuration near the crystal boundary, effectively forming a lenticular structure. 

\subsection{Cleaner bilayers with anharmonic potential}
\label{sec:anharmonic_eq}
In previous work, the addition of an anharmonic trapping potential was shown to result in single plane crystals with a more uniform areal density~\cite{Dubin2013PRA}. Motivated by this result, here we study how anharmonicity affects bilayer crystal configurations. First, we demonstrate that anharmonicity leads to a remarkable improvement in the bilayer quality. Subsequently, we provide an intuitive explanation for this effect.  

We consider the addition of an anharmonic trapping potential that modifies the potential energy~(\ref{eqn:harmonic_phi}) to $e\phi_j\rightarrow e\phi_j+e\phi_{j,\mathrm{an.}}$, where 
\begin{align}
    e\phi_{j,\mathrm{an.}} = \frac{1}{2}m\omega_z^2\frac{\beta \cfb}{r_{p,0}^2} \left[z_j^4 - 3z_j^2(x_j^2 + y_j^2) + \frac{3}{8} (x_j^2 + y_j^2)^2\right].
    \label{eqn:anharmonic_phi}
\end{align}
Here, we find it convenient to quantify the strength of the anharmonic potential by the parameter $\cfb$, which is made dimensionless by introducing a length scale $r_{p,0}$ in Eq.~(\ref{eqn:anharmonic_phi})~\footnote{We note that our definition of $\cfb$ differs slightly from that used in Ref.~\cite{Dubin2013PRA}. In that work, $\cfb$ is made dimensionless by introducing a length scale $r_p$ corresponding to the plasma radius \emph{in the presence} of an anharmonic potential of strength $\cfb$. Since $r_p$ implicitly depends on the value of $\cfb$, which makes the anharmonic potential nonlinear in $\cfb$. Here, we prefer to have a linear relation between $e\phi_{j,\mathrm{an.}}$ and $\cfb$, and hence set the length scale as $r_{p,0}$, which is independent of $\cfb$ by construction.}. The latter quantity is the radius of the crystal, called the plasma radius, in a purely harmonic trap, which can be analytically derived~\cite{Dubin2013PRA} and is given by $r_{p,0}=(3\pi N/4)^{1/3}a_0$, and $a_0=[k_e e^2/(m\omega_\perp^2)]^{1/3}$ is a length scale where the Coulomb repulsion becomes comparable to the radial trapping strength.

\begin{figure}[!h]
	\centering
	\includegraphics[width=0.65\linewidth]{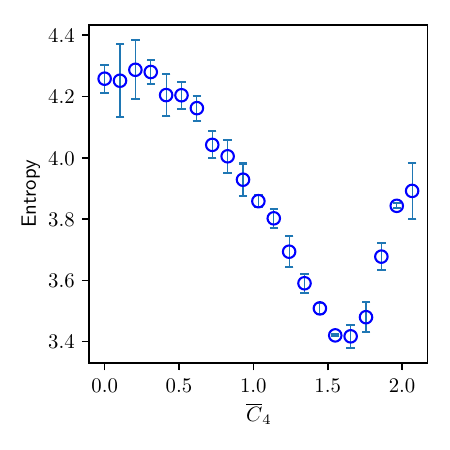}
	\caption{\textbf{Effect of anharmonic trapping on bilayer quality.} Entropy of the $z$ distribution of ions, Eq.~(\ref{eqn:entropy}), versus the strength of the anharmonic term in the trapping potential [Eq.~(\ref{eqn:anharmonic_phi})]. Here, we choose $\omega_r/(2\pi)=210$ kHz and take other trapping parameters to be the same as in Fig.~\ref{fig:harmonic_eq}. The entropy attains a minimum at an optimal value of $\cfb$, showing that bilayer quality can be improved by using anharmonic trapping potentials. The bin size used to make the histograms for computing the entropy is ten times smaller than that used for the histograms shown in Fig.~\ref{fig:harmonic_eq} and Fig.~\ref{fig:anharmonic_eq}, since finer bins lead to a more robust estimate for an optimal $\cfb$ that is less sensitive to the choice of bin edges.
}
 \label{fig:entropy}
\end{figure}

\begin{figure*}[!t]
	\centering
	\includegraphics[width=0.9\linewidth]{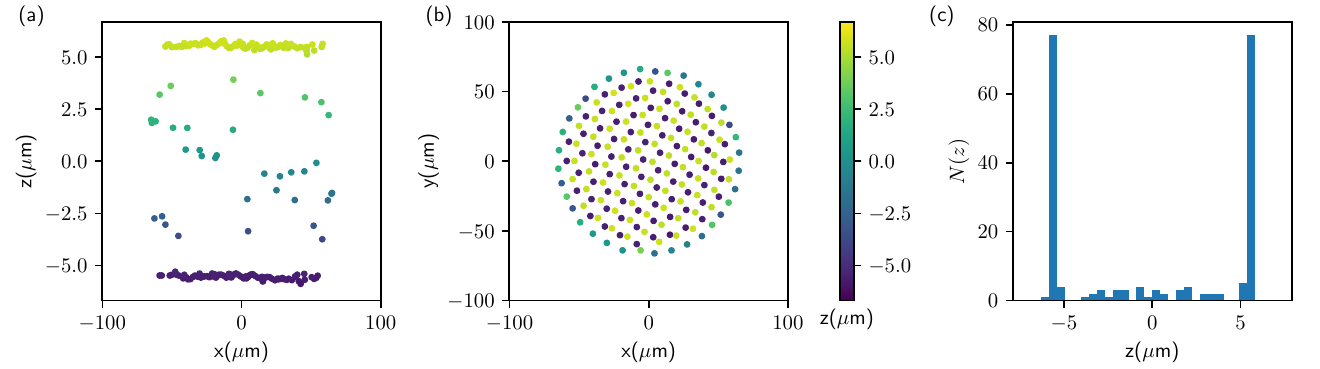}
	\caption{\textbf{Clean bilayer crystals with anharmonic trapping.} The three panels show the side view, top view and histogram of $z$ positions for a bilayer crystal obtained under anharmonic trapping with $\cfb\approx1.63$. Other trapping parameters are the same as in Fig.~\ref{fig:entropy}. The layers are now visibly planar, are well-demarcated and with minimal boundary effects, compared to the crystals in Fig.~\ref{fig:harmonic_eq}. In the rest of this work, we will study the properties of this crystal and its applications in quantum information processing. 
 }
 \label{fig:anharmonic_eq}
\end{figure*}

We consider a crystal with $N=200$ ions that has a $\beta/\beta_c=1.938$ when the trap is harmonic, i.e. $\cfb=0$, and study how the histogram of $z$ positions changes as $\cfb$ is increased. Specifically, as a quantitative measure of bilayer quality, we introduce the entropy 
\begin{align}
    S = -\sum_j f_k \ln f_k,
    \label{eqn:entropy}
\end{align}
where $f_k = N(z_k)/N$ are the normalized counts in the $k^{\rm th}$ bin with center at $z_k$. Figure~\ref{fig:entropy} shows how the entropy $S$ changes with $\cfb$. Since our global minimization routine  (Appendix~\ref{app:Modified_Basin_hopping}) is not a deterministic algorithm, for each value of $\cfb$, we generate the equilibrium configuration $50$ times and plot the value of $S$ averaged over only the lowest $10$ energy configurations, with the associated standard deviation as the error bar. The entropy decreases as $\cfb$ is increased and is minimized at an optimal value of $\cfb\approx1.65$. Further increasing $\cfb$ leads to an increase in $S$. We note that the precise location of the optimal $\cfb$ value is sensitive to the bin size and the location of the bin edges used to make the histogram, so the entropy measure should be understood as a guide to identify a ballpark value of $\cfb$ required to produce clean bilayers. In the Supplemental Material, we provide an animation of how the crystal configuration progressively improves as $\cfb$ is increased by plotting the lowest energy configurations obtained for different $\cfb$. 

Figure~\ref{fig:anharmonic_eq} shows the side and top views of a bilayer configuration obtained at $\cfb\approx 1.63$, along with the corresponding histogram of $z$ positions. The side view and histogram demonstrate the remarkable improvement in the bilayer quality. Not only are the two layers more planar as compared to the crystals in Fig.~\ref{fig:harmonic_eq}, but the number of ions in the `scaffolding' surrounding the two layers is also greatly reduced. The top view shows that the lattice structure in this configuration is still hexagonal with the two layers' lattices staggered with respect to each other.

Our study above shows that high-quality bilayers with minimal boundary effects and distortions are realizable in Penning traps by the inclusion of an anharmonic trapping potential. In Appendix~\ref{app:AnharmonicityExperimental}, we consider the practical feasibility of generating a strong anharmonic trapping potential term such that $\cfb\gtrsim 1$. Although in the NIST Penning trap, with $\omega_z/(2\pi)=1.62$ MHz and for an $N=200$ ion crystal, we find that $\cfb \lesssim 0.0035$ under realistic trap voltage constraints, our analysis in Appendix~\ref{app:AnharmonicityExperimental} indicates that it can be strongly enhanced by using larger ion numbers ($\cfb\propto N^{3/2}$), lower axial trapping frequencies ($\propto \omega_z^{-10/3}$), and smaller trap dimensions ($\propto d^{-4}$). Hence, achieving $\cfb\gtrsim 1$ is indeed experimentally viable, in principle even in the present trap at NIST. 

\subsubsection{Effect of anharmonicity: Qualitative explanation}

The role of the $\cfb$ term in producing cleaner bilayer crystals can be intuitively understood using qualitative considerations of force balance. In a purely harmonic potential Fig.~\ref{fig:harmonic_eq} shows that the two `layers' of the crystal curve inwards as a function of radial distance $\rho=\sqrt{x^2+y^2}$ from the trap center. This curvature can be understood as arising from the balance of the external trapping and the Coulomb repulsion of the two layers: Assuming each layer to be an approximately flat disk, the harmonic trapping pushes each disk inward with a force that is independent of $\rho$. However, the repulsive Coulomb force of one disk on another along $z$ is largest at the center and decreases with $\rho$. As a result, the central region of each layer tends to bulge outward along $z$, whereas they curve inward with increasing $\rho$. 

In the presence of an anharmonic term, the radial decrease in the Coulomb force can be compensated by the term $\propto -z^2\rho^2$  in Eq.~(\ref{eqn:anharmonic_phi}), which leads to an outward force on each layer that \emph{increases} with $\rho$. This effectively leads to a more uniform force along the $z$ direction over the entire layer, thereby leading to the formation of flatter layers, as seen in Fig.~\ref{fig:anharmonic_eq}. Furthermore, we observe that for values of $\cfb$ beyond the optimal range, each layer has an outward curvature indicating that the anharmonic term is now overcompensating for the radial reduction in Coulomb force (see animation of equilibrium crystal configuration versus $\cfb$ provided in the Supplemental Material). We note that a similar argument applied along the radial direction using the term $\propto \rho^4$ in Eq.~(\ref{eqn:anharmonic_phi}) can be used to intuitively understand the formation of single-plane crystals with more uniform areal density, which has been previously analyzed in  Ref.~\cite{Dubin2013PRA}.

\section{Normal Modes of Bilayer crystals}
\label{sec:normal_modes}

In order to use bilayer crystals for quantum information processing, we study the properties of their normal modes of motion. In this section, we first briefly recall the normal mode analysis for crystals in Penning traps~\cite{wang2013PRA,dubin2020POP,shankar2020PRA}. Subsequently, we compare the normal modes of bilayers and single-plane crystals using multiple metrics and show the dramatic difference in the nature of the drumhead (axial) modes in the bilayer regime. With an eye on quantum applications, we then quantize the modes using a procedure that makes the quantized modes amenable to a transparent physical interpretation. 

\subsection{Normal mode analysis}
\begin{figure*}[!tb]
    \includegraphics[width=0.95\linewidth]{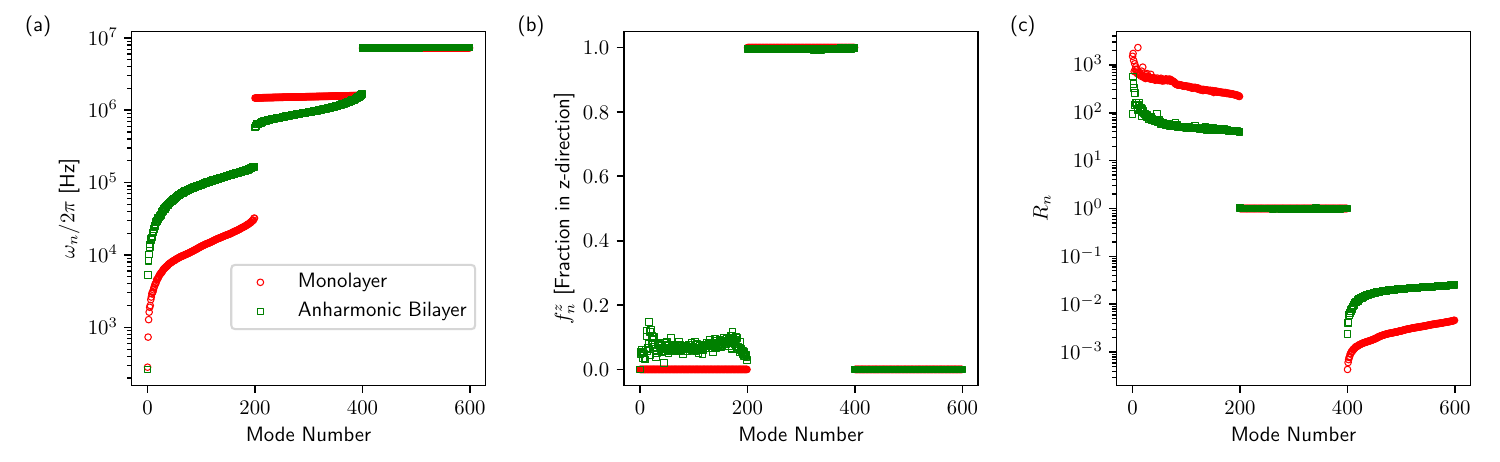}
    \caption{\textbf{Properties of normal modes of bilayer crystals.} The three panels show (a) the normal mode frequencies, (b) the fraction of the mode contributed by out-of-plane motion [Eq.~(\ref{eqn:f_nz})], and (c) the ratio of the average potential to kinetic energy in the mode [Eq.~(\ref{eqn:r_n})], for the $3N=600$ modes of the $N=200$ ion bilayer crystal shown in Fig.~\ref{fig:anharmonic_eq}. For comparison, we also plot the corresponding quantities for a single-plane crystal, which is obtained by choosing $\omega_r/(2\pi)=180$ kHz and taking other parameters to be the same as in Fig.~\ref{fig:harmonic_eq}. For both crystals, the normal modes can be classified into three branches: a low-frequency $\exb$ branch (modes $0-199$), an intermediate frequency drumhead branch ($200-399$) and a high-frequency cyclotron branch ($400-599$).
    }
    \label{fig:w_R_fz}
\end{figure*}
The normal modes of motion are obtained by studying the small-amplitude motion of the ions about the equilibrium configuration. We define a composite phase-space vector $\dKet{q}=(\dKet{\delta r},\dKet{v})^T$, where $\dKet{\delta r},\dKet{v}$ respectively denote the small-amplitude \emph{classical} position and velocity displacements of the ions \footnote{In this paper, all vectors (not quantum states) are represented as $\dKet{\cdot}$, and the corresponding inner product, outer product, and matrix element are represented as $\dip{\cdot}{\cdot}$, $\dop{\cdot}{\cdot}$, and $\dmel{\cdot}{\cdot}{\cdot}$ respectively. The quantum states are still represented using $\ket{\cdot}$ and the corresponding standard nomenclature.}. The vectors $\dKet{\delta r},\dKet{v}$ are $3N$-dimensional as they account for the $x,y,z$ degrees of freedom of all the ions. We note that, at this point, the motion is treated as classical and we are using the bra-ket notation to represent vectors and inner products purely for notational convenience. These quantities will be quantized in Sec.~\ref{sec:Quantization}, where, e.g., the corresponding vector of $3N$ position operators will be denoted with hats as $\dKet{\delta\hat{r}}$~[Eq.~(\ref{eqn:delta_r_quantized})]. These displacements can be expressed in terms of the $3N$ normal modes of the crystal as 
\begin{align}
\dKet{q} = \sum_{n=1}^{3N} \left(A_n e^{-i\omega_n t}\dKet{u_n} +  A_n^* e^{i\omega_n t}\dKet{u_n^*} \right),
\label{eqn:q_vec}
\end{align}
where $\omega_n$, $\dKet{u_n}$ and $A_n$ are respectively the eigenfrequencies, eigenvectors and complex amplitude associated with the $n$th mode. Explicitly writing the position and velocity components of the eigenvector as $\dKet{u_n} = (\dKet{u_n^r},\dKet{u_n^v})^T$, the eigenvalue equations can be obtained by linearizing the Euler-Lagrange equations and are given by
\begin{align}
    &\dKet{u_n^v} = -i\omega_n\dKet{u_n^r},\nonumber\\
    -&\mathbb{K}\dKet{u_n^r} + m\mathbb{L}\dKet{u_n^v} = -im\omega_n\dKet{u_n^v}.
    \label{eqn:eig}
\end{align} 
Here, $\mathbb{K}$ is the real, symmetric stiffness matrix, whose form is given in Appendix~\ref{app:StiffnessMatrix}. A primary difference between normal modes in rf Paul traps and Penning traps arises from the Lorentz force, which introduces a velocity-dependent force through the matrix $\mathbb{L}$~\cite{shankar2020PRA}. With the basis ordered as $\{x_1,y_1,z_1,\ldots,x_N,y_N,z_N\}$, it can be expressed as $\mathbb{L} = i\omega_c'\mathrm{diag}(M_1,M_2,\ldots,M_N)$, where $M_j$ is a $3\times3$ matrix in the basis of $x_j,y_j,z_j$ given by  
\begin{align}
    M_j = M = 
    \begin{pmatrix}
    0 & -i & 0 \\
    i & 0 & 0 \\
    0 & 0 & 0
    \end{pmatrix}.
\end{align} 
This structure of $M_j$ arises from the fact that the magnetic field is along the $z$ axis and only introduces Lorentz forces in the $x-y$ plane. 

The total energy associated with the small-amplitude motion can be expressed in terms of the normal modes as 
\begin{align}
    \mathcal{H} = \sum_{n=1}^{3N} \abs{A_n}^2\left(\dmel{u_n^r}{\mathbb{K}}{u_n^r} + m\dip{u_n^v}{u_n^v} \right).
    \label{eqn:ham_classical}
\end{align}
Equation~(\ref{eqn:ham_classical}) shows that the total energy in each mode consists of separate contributions from the position and velocity components of the eigenvector, which can respectively be identified as the time-averaged potential and kinetic energies associated with that mode.

\subsection{Bilayer vs monolayer normal modes}

\subsubsection{Monolayer crystal}

For a monolayer, the linearized Euler-Lagrange equations decouple for the in-plane ($x,y$) and out-of-plane ($z$) degrees of freedom, leading to a block-diagonal form for the stiffness matrix $\mathbb{K}=\mathrm{diag}(\mathbb{K}_{xy},\mathbb{K}_{z})$~\cite{wang2013PRA,shankar2020PRA}. Therefore, as shown in Fig.~\ref{fig:w_R_fz}(a), the normal modes consist of $2N$ in-plane modes, grouped into $N$ low-frequency $\exb$ modes ($\lesssim 100$ kHz) and $N$ high-frequency cyclotron modes ($\sim 7.2$ MHz), and $N$ out-of-plane axial or `drumhead' modes that are intermediate in frequency ($\sim 1$ MHz to $1.6$ MHz for typical NIST trapping parameters). In Fig.~\ref{fig:w_R_fz}(b), we show the fraction $f_n^z$ of the mode that is contributed by out-of-plane motion, i.e.
\begin{align}
    f_n^z = \frac{\dip{u_n^{z}}{u_n^{z}}}{\dip{u_n^r}{u_n^r}}.
    \label{eqn:f_nz}
\end{align}
As expected, $f_n^z=0$ for the $\exb$ and cyclotron mode branches, whereas $f_n^z=1$ for the drumhead modes. 

The Lorentz force has a nontrivial effect on the nature of the normal modes. In contrast to crystals in rf Paul traps, the normal modes of ion crystals in Penning traps do not in general correspond to simple harmonic motion. Consequently, the time-averaged potential and kinetic energy content in a mode may not be equal. This is quantified by the metric $R_n$~\cite{shankar2020PRA}, defined as 
\begin{align}
    R_n = \frac{\dmel{u_n^r}{\mathbb{K}}{u_n^r}}{m\omega_n^2\dip{u_n^r}{u_n^r}},
    \label{eqn:r_n}
\end{align}
which gives the ratio of the average potential to kinetic energy in the $n$th mode. As shown in Fig.~\ref{fig:w_R_fz}(c), $R_n\gg 1$ ($R_n\ll 1$) for the $\exb$ (cyclotron) branches. In the case of the drumhead modes, $R_n=1$ identically, since these modes decouple from the in-plane motion and are simple harmonic in nature, i.e. they are solutions to a system of coupled simple harmonic oscillators and satisfy 
\begin{align}
    \mathbb{K}_z\dKet{u_n^{z}}=m\omega_n^2\dKet{u_n^{z}}.
    \label{eqn:mono_drum}
\end{align}

\subsubsection{Bilayer crystal}

On the other hand, the linearized Euler-Lagrange equations for the in-plane and out-of-plane motion do not decouple for bilayer crystals. In Fig.~\ref{fig:w_R_fz}(a), we plot the eigenfrequencies of the $3N$ normal modes for the clean anharmonic bilayer shown in Fig.~\ref{fig:anharmonic_eq}, and find that, similar to the monolayer, there are three distinct branches of low, intermediate and high frequency modes. Hence, we will refer to these branches using the same terminology as for the monolayer modes, viz., $\exb$, drumhead and cyclotron modes.  Compared to the monolayer, the $\exb$ modes in general have higher frequencies in the bilayer regime, with the exception of the lowest frequency `rocking' mode~\cite{tang2019POP}, which shifts to a lower frequency. The drumhead modes now have an increased bandwidth, and hence the separation between the $\exb$ and drumhead branches is smaller. On the other hand, there is no significant difference in the frequencies of the high-frequency cyclotron modes, on the scale shown here. 

An important feature of the $\exb$ modes in the bilayer crystal is that they now have a non-negligible fraction $f_n^z\sim 0.05-0.1$ of out-of-plane motion. This implies that, with bilayer crystals, the $\exb$ modes can be addressed using standard setups for quantum information processing in Penning traps, where lasers are used to couple the electronic states of ions to their out-of-plane motion in a frequency-selective manner. Such a capability could be used, e.g. for improved laser cooling and thermometry of the $\exb$ modes and in using these modes for engineering interacting spin models for quantum simulation. On the other hand, the $f_n^z$ values for the drumhead and cyclotron modes do not change significantly compared to the monolayer, but they are no longer identically equal to $1$ and $0$ respectively.  

Similar to the monolayer, the $\exb$ (cyclotron) mode branches are dominated by potential (kinetic) energy, although the deviation of $R_n$ from 1 is reduced for the bilayer crystal. The drumhead modes are no longer simple harmonic and are instead obtained as solutions to the full set of equations~(\ref{eqn:eig}). Nevertheless, they are still found to have $R_n\approx 1$, i.e., they have nearly equal average potential and kinetic energies. 

\subsubsection{Complex drumhead modes in bilayer crystals}

All the metrics considered so far, viz., the mode frequencies, $f_n^z$ and $R_n$ indicate that the drumhead modes of a bilayer crystal are qualitatively similar to those of a monolayer and only differ quantitatively. Furthermore, the differences in metrics such as $f_n^z$ and $R_n$ are barely noticeable, since they only deviate very slightly from $1$. However, these deviations point to the coupled nature of the out-of-plane and in-plane motions in the bilayer case, which leads to a dramatic change in the nature of the \emph{eigenvectors} of the drumhead modes.

\begin{figure}
    \includegraphics[width=0.6\linewidth]{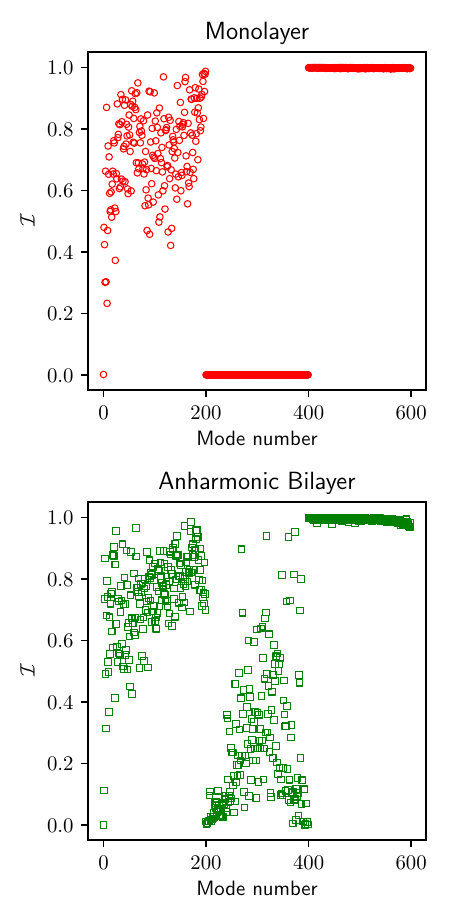}
    \caption{\textbf{Complex eigenvectors of drumhead normal modes,} The two panels plot $\mathcal{I}_n$ [Eq.~(\ref{eqn:imaginarity})] versus mode number $n$, for a single-plane crystal (top panel) and for a bilayer crystal, with the same parameters as in Fig.~\ref{fig:w_R_fz}. The central difference is the emergence of complex eigenvectors in the drumhead branch (mode numbers $200-399$) in the bilayer crystal. In contrast, all drumhead modes have real eigenvectors in the single-plane crystal.}
    \label{fig:mode_complexity}
\end{figure}

In the monolayer, the drumhead modes satisfy Eq.~(\ref{eqn:mono_drum}) and hence the eigenvectors $\dKet{u_n^r}$ are all real. However, this is not generally true for an eigenvector satisfying Eq.~(\ref{eqn:eig}). As a measure of the mode complexity, we compute the metric
\begin{align}
\mathcal{I}_n = \frac{\abs{\dip{u_n^r}{u_n^r}}-\abs{\dip{u_n^r}{u_n^{r*}}}}{\abs{\dip{u_n^r}{u_n^r}}+\abs{\dip{u_n^r}{u_n^{r*}}}},
\label{eqn:imaginarity}
\end{align}
which is identically $0$ for real eigenvectors, as occurs, e.g. for simple harmonic motion, and can have a maximum value of $1$, as occurs, e.g. for perfect circular motion. In Fig.~\ref{fig:mode_complexity}, we plot $\mathcal{I}_n$ for all the modes of a monolayer as well as a bilayer crystal. While the $\exb$ and cyclotron branches generally have non-zero $\mathcal{I}_n$ values in both cases, the remarkable feature is the emergence of complex eigenvectors in the drumhead branch (mode numbers $201-400$) in the bilayer crystal. The degree of complexity is appreciable despite the drumhead modes still being predominantly out-of-plane in nature, i.e. $f_n^z\approx 1$. From animations of drumhead mode motion (see Supplemental Material), we observe that although the ion displacements are mostly along the $z$ direction, the complex eigenvectors arise from a chiral propagation of the disturbance associated with these modes in the bilayer crystal. In other words, nontrivial phase relations are established between the out-of-plane displacements of different ions. In contrast, no such chiral propagation is observed for drumhead modes of monolayer crystals, where, as the purely real eigenvectors imply, the ions only move either in phase or out-of-phase with respect to one another.  

By supporting drumhead modes with complex eigenvectors, bilayer crystals in Penning traps provide a feature that is not natively present in rf Paul traps or monolayer Penning trap crystals, where all the normal modes used for quantum information processing are associated with purely real eigenvectors. This opens up a new potential pathway to engineer chiral interactions, which occur, e.g., in spin models with a Dzyaloshinskii–Moriya (DM) interaction~\cite{dzyaloshinsky1958JPhysChemSol,moriya1960PR}. 

\subsubsection{Some drumhead modes of interest}

\begin{figure*}
    \includegraphics[width=0.99\linewidth,trim={0 0 0 1cm},clip]{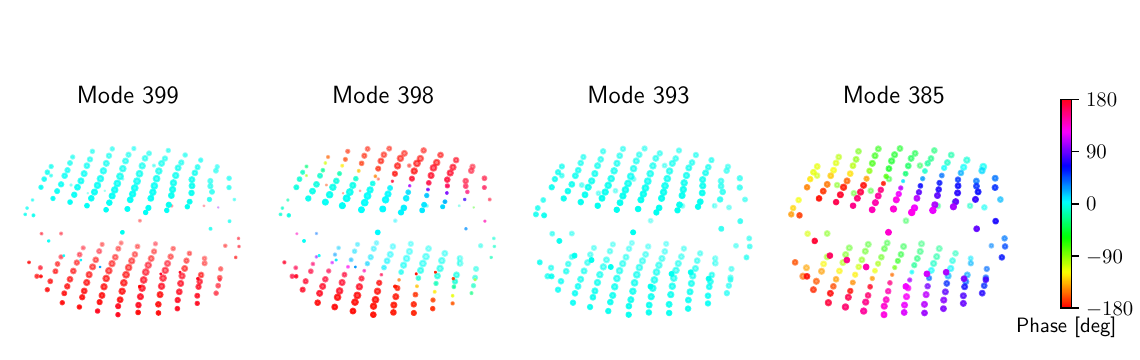}
    \caption{\textbf{Representative drumhead normal modes of bilayer crystals.} Eigenvectors of four different drumhead modes, viz., the breathing mode (mode number $399$), a bilayer tilt mode ($398$), the center-of-mass mode ($393$) and a chiral mode with $\mathcal{I}_n\approx 0.8$ ($385$). The eigenvectors are in general complex, and hence we use the size of the ion marker to indicate the amplitude of the ion participation in a mode, and the color of the marker to represent the phase of the ion's motion. These modes correspond to the crystal shown in Fig.~\ref{fig:anharmonic_eq}.}
    \label{fig:anharm_modes}
\end{figure*}

Quantum simulation and sensing applications enabled by bilayer crystals will rely on coupling the electronic states of the ions to the drumhead modes. Hence, it is useful to gain intuition into the nature of these modes. In Fig.~\ref{fig:anharm_modes}, we illustrate the eigenvectors of four such modes for the anharmonic bilayer shown in Fig.~\ref{fig:anharmonic_eq}. For each mode, the size of the marker for each ion is proportional to the mode amplitude ($\abs{u_{n,j}^z}$) supported by that ion, while the color represents the phase [$\mathrm{arg}(u_{n,j}^z)$] of motion. 

For this bilayer crystal, mode number $393$ corresponds to the center-of-mass (c.m.) mode, where all the ions move in phase. However, because of the anharmonicity, all the ions do not have equal amplitude, with ions at the boundary (center) having smaller (larger) amplitudes. Furthermore, unlike in a monolayer, the c.m. mode is not the highest-frequency drumhead mode, which is instead found to be a breathing mode (mode  $399$). In this mode, the ions in the top and bottom layers move out-of-phase with respect to each other and the amplitude of motion decreases as the distance from the center increases. This mode can be viewed approximately as two monolayer c.m. modes with a $180^\circ$ phase shift between the two layers. In fact, several high-frequency drumhead modes in the bilayer are actually generalizations of well-known monolayer drumhead modes with a $180^\circ$ phase shift between the two layers. This feature is very evident in the pair of bilayer tilt modes, one of which (mode $398$) is shown here. The out-of-phase tilting motion is reflected in the color maps used to represent the phase here. 

The three modes discussed so far, viz., c.m., breathing and tilt modes have predominantly real eigenvectors, leading to $\mathcal{I}_n \lesssim 10^{-2}$. This is reflected in the corresponding phase maps, which predominantly have values corresponding to only $0^\circ$ or $180^\circ$. As a final example, we consider mode $385$, which has a \emph{complex} eigenvector with $\mathcal{I}_n\approx 0.8$, and hence represents a qualitatively different kind of drumhead mode that is not found in a monolayer crystal. In this mode, both the layers undergo tilting motion with a common tilt axis that is not fixed, but is instead rotating with time. We provide an animation in the Supplemental Material. Remarkably, the phase of motion displays a clockwise circulation ($0^\circ\to90^\circ\to180^\circ$) in each layer, as shown in Fig.~\ref{fig:anharm_modes}. In fact, this mode is a bilayer analog of an $(l=2, m=1)$  electrostatic fluid mode occuring in spheroidal non-neutral plasmas~\cite{dubin1991PRL,bollinger1993PRA}. Here, $l$ and $m$ characterize the mode in terms of an associated Legendre polynomial  $P_l^m(\xi_2)$, where $\xi_2$ is a generalized `latitude' (polar angle) coordinate, and an $e^{im\varphi}$ dependence on the azimuthal angle $\varphi$ corresponding to the ion position. This mode is accompanied by a corresponding mode with counterclockwise circulation (also shown in the animation), which is the bilayer analog of an $(l=2, m=-1)$ mode of a spheroidal plasma.

\subsection{Quantization of normal modes}\label{sec:Quantization}

A quantum description of the normal modes is required in order to study their use in quantum information processing. For a monolayer, the quantization of the drumhead modes is straightforward since they are simple harmonic oscillator modes. The quantization of the $\exb$ and cyclotron modes was considered in Ref.~\cite{wang2013PRA} and was performed using an involved procedure. Here, we present a unified method to quantize the normal modes of a general single-species 3D crystal in a Penning trap, using steps that make the physical content of the quantization procedure transparent.

Our method involves substituting in Eq.~(\ref{eqn:q_vec}) an ansatz for the complex mode amplitudes $A_n \to c_n\hat{a}_n, A_n^* \to c_n\hat{a}_n^\dag $, where $c_n$ are real coefficients discussed below and $\hat{a}_n,\hat{a}_n^\dag$ are annihilation and creation operators. Subsequently, we prove that if these operators satisfy the usual bosonic commutation relations 
\begin{align}
 [\hat{a}_n,\hat{a}_{n'}^\dag]=\delta_{n,n'},   
 \label{eqn:a_adag_comm}
\end{align}
then the canonical commutation relation for the position and canonical momentum of every ion is automatically satisfied, i.e. 
\begin{align}
    [\hat{x}_j,\hat{p}_{x,j}]=[\hat{y}_j,\hat{p}_{y,j}]=[\hat{z}_j,\hat{p}_{z,j}]=i\hbar, \; \forall j.
    \label{eqn:canonical_comm}
\end{align}
Here, $\hat{\alpha}_j, \alpha=x,y,z$ are position operators corresponding to displacements of ion $j$ from its equilibrium position, and $\hat{p}_{\alpha,j}, \alpha=x,y,z$ are the operators corresponding to the canonical momentum $p_{\alpha,j}=\partial L/\partial \dot{\alpha}_j$, which is in general different from the mechanical momentum $\Pi_{\alpha,j}=m\dot{\alpha}_j$ because of the magnetic field. The details of the proof are presented in Appendix~\ref{app:ModeQuantization}. Here, we present a brief summary of the central steps. First, the $3N$ commutators in Eq.~(\ref{eqn:canonical_comm}) are expanded in terms of the normal mode operators and simplified using the corresponding commutation relation~(\ref{eqn:a_adag_comm}). Subsequently, the expressions reduce to sums over terms involving the normal mode frequencies and eigenvectors alone, which must be shown to equal $1$ in order to satisfy~(\ref{eqn:canonical_comm}). Intuitively, these $3N$ constraints on the normal modes arise due to the equivalence in expressing the total energy per degree of freedom of each ion (i) in terms of the ion's displacement and momentum, or (ii) in terms of its contribution to the different normal modes. In Appendix~\ref{app:ModeQuantization}, we rigorously prove that these constraints are satisfied by applying the equipartition theorem.  

Using the quantized normal modes, the vector of $3N$ operators representing position fluctuations can be expressed as
\begin{align}\label{eqn:delta_r_quantized}
    \dKet{\delta\hat{r}} = \sum_n c_n \left(\hat{a}_n e^{-i\omega_n t}\dKet{u_n^r} + \hat{a}_n^\dag e^{i\omega_n t}\dKet{u_n^{r*}}\right),
\end{align}
where the coefficients $c_n=l_{0,n}/\sqrt{\dip{u_n^r}{u_n^r}}$ and $l_{0,n}$ is a length scale that represents the root-mean-square (RMS) zero-point fluctuation of the $n$th mode, i.e. it is the total RMS fluctuation of all the ions in the crystal arising because of mode $n$. It is given by 
\begin{align}
    l_{0,n}^2 = \frac{\hbar}{m\omega_n(1+R_n)},
    \label{eqn:l0}
\end{align}
where $R_n$ is given by Eq.~(\ref{eqn:r_n}). We note that, for any drumhead mode of a monolayer, $R_n=1$ and the expression reduces to the familiar form $\hbar/(2m\omega_n)$ for the zero-point fluctuations of a simple harmonic oscillator.

Furthermore, the quantized Hamiltonian corresponding to Eq.~(\ref{eqn:ham_classical}) can be shown to reduce to the expected form 
\begin{align}
    \hat{H} = \sum_{n} \hbar\omega_n\left(\hat{a}_n^\dag\hat{a}_n+\frac{1}{2} \right).
    \label{eqn:ham_quantized}
\end{align}
From Eqs.~(\ref{eqn:delta_r_quantized}) and~(\ref{eqn:ham_quantized}), it can be seen that $\dKet{\delta\hat{r}}$ corresponds to the position fluctuations in an interaction picture taken with respect to the Hamiltonian corresponding to the free evolution of the normal modes.

\section{Quantum control of Bilayer crystals}
\label{sec:quantum_control}
\begin{figure}
    \centering
    \includegraphics[width=0.9\columnwidth]{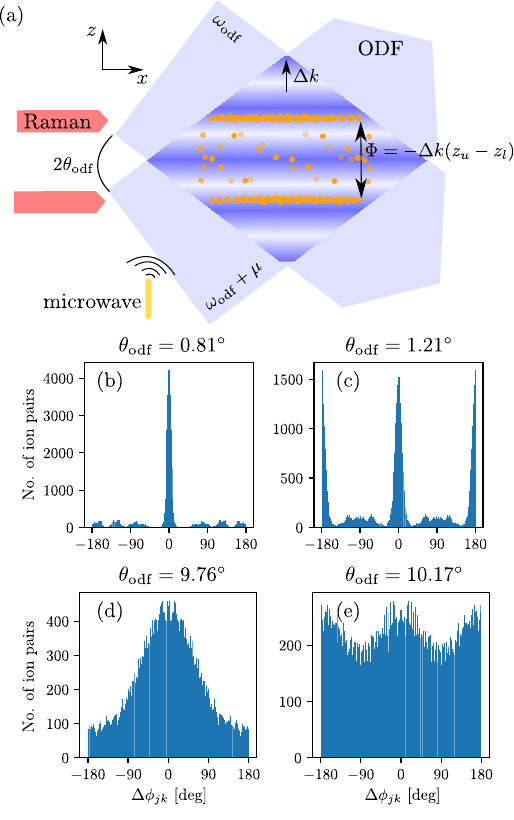}
    \caption{\textbf{Quantum control of bilayer crystals.} (a) Single spin and entangling operations: Microwaves and Raman beams can be respectively used for implementing global and layer-selective single spin (ion) rotations. Spin-spin entanglement is enabled by an optical dipole force (ODF) that is generated by a pair of lasers intersecting the crystal at angles $\pm\theta_\mathrm{odf}$ to the $x-y$ plane. Their interference creates a moving optical lattice that leads to spin-motion coupling, which in turn mediates effective spin-spin interactions. Bilayer crystals present a new control knob, viz., the mean interlayer ODF phase difference $\Phi$ [Eq.~(\ref{eqn:mean_phase_diff})], which is a function of $\theta_\mathrm{odf}$ and can be used to tune the relative interlayer to intralayer couplings. Panels (b)-(e): Histogram of the ODF phase difference $\Delta\phi_{jk}$ [Eq.~(\ref{eqn:phase_diff})] between all pairs of ions $j,k$ in the crystal shown in Fig.~\ref{fig:anharmonic_eq}, and for different values of $\theta_\mathrm{odf}$. Sharp peaks in the histograms in (b) and (c) show that each layer effectively appears as a near-perfect plane at grazing incidence of the ODF lasers ($\theta_\mathrm{odf}\sim 1^\circ$).}
    \label{fig:setup}
\end{figure}

In this section, we set the stage for quantum information processing with bilayer crystals by identifying the resources available for quantum control on this platform. After a brief description of single spin operations, we model the application of the optical dipole force (ODF) for entangling ions in bilayer crystals. We show how the analysis of the ODF interaction provides a natural route to assessing the quality of the bilayer.

\subsection{Single spin operations}

A spin-$1/2$ system, described by Pauli operators $\hSig_j^\alpha,\alpha=x,y,z$, is encoded in each ion $j$ by utilizing the two long-lived hyperfine states in the ground ${}^2\mathrm{S}_{1/2}$ manifold of ${}^9\mathrm{Be}^+$ as the $\ket{\uparrow},\ket{\downarrow}$ states~\cite{biercuk2009QIC}. As illustrated in Fig.~\ref{fig:setup}, microwave driving of the two hyperfine states is typically used to perform identical single spin operations on all ions in the crystal. Additionally, in bilayer crystals, the large separation between the two layers ($\sim 10 \;\mu$m, see Fig.~\ref{fig:anharmonic_eq}) provides an opportunity to perform layer-selective single-spin operations by addressing the two layers with separate pairs of Raman beams from the side of the crystal~\footnote{We note that because of Doppler shifts arising from the crystal rotation, the Raman Rabi frequency will slightly decrease with the ion distance from the trap center. For crystals with radius in the range $75\;\mu$m to $100\;\mu$m, the fractional decrease in Rabi frequency from the trap center to the crystal boundary is less than $2\%$. This is discussed in more detail in Appendix B of Ref.~\cite{shankar2022PRXQ}.}. With the use of cylindrical optical elements, these beams can be engineered to have an elliptical beam waist, such that the waist along the $z$ direction is a few microns while the waist along the $y$ direction is several tens to hundreds of microns in order to cover the planar extent of each layer.

The spatial separation of the two layers also allows for performing layer-resolved readout of the spin states of the ions via detection of fluorescence from the side of the crystal.

\subsection{Entangling resource: Optical Dipole Force}\label{sec:ODF}

The normal modes of motion, typically the drumhead modes, serve as the quantum channels that couple the spins and enable entanglement generation~\cite{britton2012nat}. The required spin-motion coupling is generated via an optical dipole force (ODF) that is implemented using either a light-shift gate or a M\o lmer-S\o rensen (MS) gate~\cite{carter2023PRA}. Here, we will focus on the ODF implemented using a light-shift gate applied to bilayer crystals, and briefly comment on the MS gate in Sec.~\ref{sec:conclusion}. 

As shown in Fig.~\ref{fig:setup}, the ODF is applied using a pair of lasers with frequencies $\omega_\mathrm{odf},\omega_\mathrm{odf}+\mu_r$ that are incident on the crystal at angles of $\pm\theta_\mathrm{odf}$ with respect to the $x-y$ plane. The interference of these two beams results in a traveling wave optical lattice along the direction of their difference wavevector $\Delta k\approx 2 k_\mathrm{odf}\sin\theta_\mathrm{odf}$, which in this configuration is along the $z$ direction. Here, $k_\mathrm{odf}$ is the wavevector magnitude, which is approximately the same for each laser. This lattice results in the ions experiencing a spatially varying ac Stark shift~\cite{sawyer2012PRL}, which is described by the Hamiltonian 
\begin{align}
    \hat{H}_\mathrm{odf} = \sum_{j=1}^N \frac{F_0}{\Delta k} \sin(\Delta k(z_{0,j}+\hat{z}_j)-\mu_r t)\hSig_j^z,
    \label{eqn:hodf_1}
\end{align}
where $F_0$ is the magnitude of the ODF, and the $z$ position of the $j$th ion is decomposed into its equilibrium position $z_{0,j}$ and the displacement from this equilibrium described by the operator $\hat{z}_j$. For small-amplitude displacement, i.e. $\Delta k \ev{\hat{z}_j^2}^{1/2}\ll 1$, $\hat{H}_\mathrm{odf}$ can be approximated as 
\begin{align}
    \hat{H}_\mathrm{odf} \approx \sum_{j=1}^N F_0 \cos(\mu_r t+\phi_j) \hat{z}_j \hSig_j^z,
    \label{eqn:hodf_2}
\end{align}
where $\phi_j=-\Delta k z_{0,j}$. Here, we have neglected the leading-order term proportional to $\sin(\mu_r t+\phi_j)\hSig_j^z$. This term does not contain the position operator, and hence leads to a position-independent but time-varying ac Stark shift on the ions. However, this Stark shift is rapidly oscillating since $\mu_r$ is typically chosen close to the frequency of a drumhead mode, and hence this term can be neglected.

The phase difference 
\begin{align}
    \Delta\phi_{jk}=\phi_j-\phi_k
    \label{eqn:phase_diff}
\end{align} 
is a key quantity that strongly affects the interaction of pairs of ions $j,k$ via the ODF. This quantity is trivially zero for a monolayer, whereas it provides a new control knob in bilayer crystals. While the mean intralayer phase difference is zero, the mean interlayer phase difference
\begin{align}
    \Phi = -\Delta k (z_u-z_l)\,\text{mod}\,2\pi,
    \label{eqn:mean_phase_diff}
\end{align}
can be controlled by tuning the ODF angle $\theta_\mathrm{odf}$. Here, $z_u,z_l$ are the mean $z$ positions of ions in the upper and lower layers respectively. In a similar spirit, we note that experiments using linear strings of a small number of ions have demonstrated the ability to tune the ODF phase at the location of individual ions by adjusting the spacing of the ions through changes in the trap frequency~\cite{rowe2001Nat,seck2020NJP}.
In Sec.~\ref{sec:prospects}, we will demonstrate how tuning $\Phi$ provides a way to control the relative strength and phase of interlayer to intralayer interactions in this system.

\subsection{Assessing Bilayer Quality}

Although visual inspection of crystals such as in Fig.~\ref{fig:anharmonic_eq} suggests that clean bilayer crystals are produced, an operationally meaningful way to assess bilayer quality is to examine how well the ODF lattice resolves the individual layers. In Fig.~\ref{fig:setup}(b-e) we plot the histogram of phase differences $\Delta\phi_{jk}$ for \emph{all} pairs of ions $j,k$ in the crystal shown in Fig.~\ref{fig:anharmonic_eq}, and for different values of $\theta_\mathrm{odf}$. We consider two cases, viz., where $\theta_\mathrm{odf}$ is small [ $\sim 1^\circ$, Fig.~\ref{fig:setup}(b-c)] and where it is larger [ $\sim 10^\circ$, Fig.~\ref{fig:setup}(d-e)]. In turn, for each case, two panels are shown, which correspond to the  mean interlayer phase difference $\Phi = 0^\circ$ and $180^\circ$, obtained by fine tuning of $\theta_\mathrm{odf}$. For $\theta_\mathrm{odf}\sim1^\circ$, the histograms are sharply peaked, implying that the thickness of each layer is small compared to the effective wavelength of the ODF lattice, $\lambda_\mathrm{odf}=2\pi/(\Delta k)=\pi/(k_\mathrm{odf}\sin\theta_\mathrm{odf})$. On the other hand, for $\theta_\mathrm{odf}\sim10^\circ$, $\lambda_\mathrm{odf}$ is ten times smaller and the thickness of each layer spans several multiples of this wavelength, leading to a broad distribution of intralayer as well as interlayer phase differences. Indeed, the histograms of $z$ positions shown in Figs.~\ref{fig:harmonic_eq}(c), \ref{fig:harmonic_eq}(f) and~\ref{fig:anharmonic_eq}(c) are made by assuming $\theta_\mathrm{odf}=1^\circ$ and using a bin size of $\lambda_\mathrm{odf}/20$. In particular, each layer in Fig.~\ref{fig:anharmonic_eq}(c) is spread over approximately only one bin, showing that the layer thickness is only around $\lambda_\mathrm{odf}/20$ for the anharmonic bilayer crystal in Fig.~\ref{fig:anharmonic_eq}. 

The above analysis implies that operating the ODF lasers at grazing incidence to the $x-y$ plane ensures that the crystal can be approximated as a clean bilayer system to a very good extent. Hence, we will use $\theta_\mathrm{odf}\sim1^\circ$ in the following to demonstrate applications of bilayer crystals.

\section{Prospects for Quantum Information Processing}
\label{sec:prospects}

We now turn to an illustration of some of the capabilities offered by bilayer crystals in the context of quantum information processing. We first show how to generate tunable bilayer Ising models by controlling the ODF incidence angle. Subsequently, we discuss how the interlayer to intralayer coupling strength can be dynamically tuned by simultaneous coupling to two normal modes. Later, we present a route to engineer chiral spin-exchange models by the addition of a transverse field. We also discuss a number of potential applications enabled by these capabilities.

\subsection{Tunable Bilayer Ising Models}
\label{sec:tunable_ising}
\subsubsection{Control via ODF Angle}\label{sec:tunable_ising_control}

As a first example, we consider the ODF interaction when the difference frequency $\mu_r$ is tuned close to the c.m. mode frequency $\omega_\mathrm{cm}$, so that the effect of other modes can be neglected (see Sec.~\ref{sec:FrequencyResolution} and App.~\ref{app:ising_derivation}). The unitary operator for evolution under $\hat{H}_\mathrm{odf}$ for any time $t$ can be expressed as a product of a spin-motion unitary and an effective phonon-mediated spin-spin unitary~\cite{dylewsky2016PRA}. At specific `decoupling' times $\tau$ where $\abs{\delta} \tau = 2n\pi,n=1,2,\ldots$, with $\delta=\mu_r-\omega_\mathrm{cm}$, the spin-motion unitary returns to the identity operator and only the spin-spin unitary acts on the system. At these times, the state of the system can be written as $\ket{\psi(\tau)}=\hat{U}_\mathrm{ss}(\tau)\ket{\psi(0)}$, where
\begin{align}
    \hat{U}_\mathrm{ss}(\tau) = \exp\left(-i\tau\sum_{j\neq k} J_{jk}^{zz}\hSig_j^z\hSig_k^z\right)
\end{align}
is the spin-spin unitary and the exact expression for $J_{jk}^{zz}$ is given in Appendix~\ref{app:ising_derivation}. For our present discussion, the crucial observation is that coupling to the c.m. mode gives to excellent approximation, 
\begin{align}
    J_{jk}^{zz} \propto C_{jk} \cos(\Delta \phi_{jk}),
    \label{eqn:jjk_delphi}
\end{align}
where $C_{jk}$ is a positive constant and $\Delta\phi_{jk}$ is the phase difference defined in Eq.~(\ref{eqn:phase_diff}). Therefore, the interlayer coupling can be tuned in magnitude and sign by adjusting the ODF angle $\theta_{\mathrm{odf}}$, which is the knob that controls the mean interlayer phase difference $\Phi$, Eq.~(\ref{eqn:mean_phase_diff}). As a result, the interlayer coupling can be tuned from anti-ferromagnetic ($J_{jk}>0$) to ferromagnetic ($J_{jk}<0$) and also effectively turned off ($J_{jk}\approx0$) by changing $\theta_\mathrm{odf}$.

\begin{figure}
    \centering
    \includegraphics[width=0.8\columnwidth]{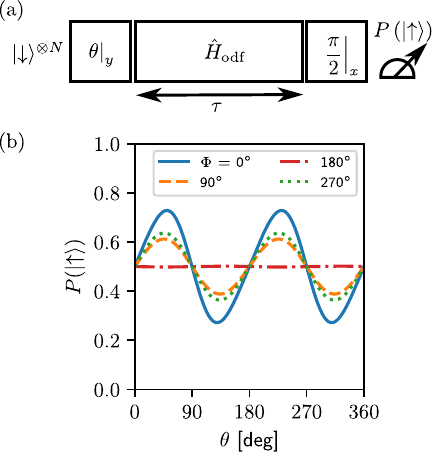}
    \caption{\textbf{Tuning interlayer Ising interactions via ODF incidence angle.} (a) Protocol to measure spin precession induced by Ising interactions as a function of the initial rotation angle $\theta$. (b) Fraction of total ions (in both layers) in $\ket{\uparrow}$ at the end of this protocol versus $\theta$, for different values of the interlayer phase difference $\Phi$ [Eq.~(\ref{eqn:mean_phase_diff})], which in turn is set by $\theta_\mathrm{odf}$. In particular, the four values of $\Phi$ listed can respectively be realized by setting $\theta_\mathrm{odf}\approx0.81^\circ, 0.61^\circ, 1.21^\circ, 1.01^\circ$ for $\Phi = 0^\circ, 90^\circ, 180^\circ, 270^\circ$. Here, we use the crystal shown in Fig.~\ref{fig:anharmonic_eq}, ODF magnitude $F_0=2.4\times10^{-23}$ N,  evolution time $\tau=750\;\mu$s, and tune the ODF close to the c.m. mode, such that $\delta/(2\pi)=1/\tau\approx 1.33$ kHz.    
    }
    \label{fig:tipping}
\end{figure}

We explore these tunable bilayer Ising models using a protocol inspired from Ref.~\cite{britton2012nat} and depicted in Fig.~\ref{fig:tipping}(a). Starting in $\ket{\downarrow}$, the spin state of all the ions is initialized at different angles in the $x-z$ plane by a global rotation about the $y$ axis by a variable angle $\theta$. Subsequently, the system is evolved under the ODF interaction for a time $\tau$. Finally, the population in the $\ket{\uparrow}$ is measured after rotating the state by $90^\circ$ about the $x$ axis.

For $4J\tau\ll 1/\sqrt{N}$, where $J$ is a typical value of $J_{jk}$, the dynamics of each spin can be understood as a precession around the $z$ axis induced by the mean field established by the $N-1$ spins. The final rotation converts the accumulated phase into a population difference between the $\ket{\uparrow}$ and $\ket{\downarrow}$ states. The probability to find ion $j$ in $\ket{\uparrow}$ at the end of the sequence is given by 
\begin{align}\label{eqn:Pj_mean_field}
    P_{j}(\ket{\uparrow}) \approx \frac{1}{2}\left[1 + \sin\left(\tau B_j \cos\theta \right)\sin\theta \right],
\end{align}
where $B_j\cos\theta$ is the precession frequency for ion $j$ with $B_j = 2\sum_{k,k\neq j}(J_{kj}+J_{jk})$. The expression~(\ref{eqn:Pj_mean_field}) is derived in Appendix~\ref{app:tipping}. The mean-field established by the $N-1$ spins is evident in the form of the precession frequency. Furthermore, since the dynamics is mediated by the c.m. mode, $B_j$ has the approximate dependence
\begin{align}
B_j\propto(1+\cos\Phi).    
\end{align}

In Fig.~\ref{fig:tipping}(b), we plot the average probability, $P(\ket{\uparrow})=\sum_j P_{j}(\ket{\uparrow})/N$, to find the ions in $\ket{\uparrow}$ versus the initialization angle $\theta$ for different ODF angles $\theta_\mathrm{odf}$, which are characterized here by the value of $\Phi$ that they establish. Although we have described a mean-field picture above to facilitate a qualitative discussion, the curves in Fig.~\ref{fig:tipping}(b) have been computed using exact expressions provided in Appendix~\ref{app:Ising}. For $\Phi=0^\circ$, the intralayer and interlayer $J_{jk}$ coefficients are positive, and pairs of ions interact in essentially the same manner regardless of which layer the two ions belong to. This leads to a strong modulation of $P(\ket{\uparrow})$ as a function of $\theta$. In contrast, for $\Phi=180^\circ$, the interlayer coefficients are negative. As a result, a net cancellation of intralayer and interlayer couplings occurs in the expression for the $B_j$ of each ion, leading to strong suppression of the mean-field dynamics. In the case of $\Phi=90^\circ,270^\circ$, the interlayer coupling is essentially turned off, and ions in one layer barely couple to ions in the other layer. This results in a reduced precession of the ion spins compared to the $\Phi=0^\circ$ case. 

\subsubsection{Dynamic Control via Two-Tone ODF}
\label{sec:dynamic}

The tunability offered by control over $\theta_\mathrm{odf}$ is attractive, but at the same time, requires mechanical adjustment between experiments to adjust the incidence angle. Here, we describe a technique to optically tune the relative strength of interlayer to intralayer coupling on-the-fly by exploiting the normal mode spectrum of bilayer crystals. Specifically, we propose to modulate one of the ODF lasers in order to produce two tones at $\omega_\mathrm{odf}+\mu_{r,0}$ and $\omega_\mathrm{odf}+\mu_{r,1}$. Interference with the other ODF laser, which is maintained at $\omega_\mathrm{odf}$ as before, now leads to two optical lattices, each with effective wavelength $\Delta k$ but with different beat frequencies $\mu_{r,0},\mu_{r,1}$. The resulting ODF Hamiltonian is given by 
\begin{align}\label{eqn:ham_two_odf}
    \hat{H}_\mathrm{odf} \approx \sum_{j=1}^N \left[F_0 \cos(\mu_{r,0} t+\phi_j) + F_1 \cos(\mu_{r,1} t+\phi_j)\right] \hat{z}_j \hSig_j^z, 
\end{align}
where $F_0,F_1$ are the effective forces induced by the two ODF lattices \footnote{ The mutual interference of the two tones leads to negligible spin-motion coupling because they propagate along the same path and hence their difference wavevector is negligible. Furthermore, with certain laser beam configurations, this interference term can be canceled out by appropriate choice of laser parameters. In addition, there can be in general be a non-zero relative phase between the two ODF lattices in Eq.~(\ref{eqn:ham_two_odf}), but this does not modify our results and hence we omit it for simplicity. We include it in our analysis in Appendix~\ref{app:MultimodeTuning}}. In this setting, we propose to tune $\mu_{r,0}$ and $\mu_{r,1}$ close to the c.m. and breathing modes respectively, such that $\abs{\delta_0}=\abs{\delta_1}=\delta$, where $\delta_0 = \mu_{r,0}-\omega_\mathrm{cm}$ and $\delta_1 = \mu_{r,1}-\omega_\mathrm{bre}$. This ensures that the spin-motion couplings induced by the c.m. mode and the breathing mode decouple at the exact same times. Furthermore, since $\omega_\mathrm{bre}-\omega_\mathrm{cm}\gtrsim 2\pi\times 100$ kHz, and typical values of $\delta\sim 2\pi \times 1$ kHz, the effects of the two ODF lattices can be independently evaluated, accounting for only the c.m. (breathing) mode for the first (second) lattice (see Appendix~\ref{app:MultimodeTuning} for a detailed discussion). For the crystal shown in Fig.~\ref{fig:anharmonic_eq}, there are only five modes lying between the breathing and c.m. modes, and their effects can be neglected in a first approximation since they are all well separated by tens of kilohertz from these two modes. The resulting spin-spin coefficient at a decoupling time is of the form
\begin{align}
    J_{jk,\mathrm{tot}}^{zz} &= J_{jk,\mathrm{cm}}^{zz} + J_{jk,\mathrm{bre}}^{zz},  \nonumber\\
    &\propto \frac{F_0^2}{\delta_0}C_{jk} + \frac{F_1^2}{\delta_1}D_{jk}, \label{eqn:Jij_tunable}
\end{align}
where $C_{jk},D_{jk}$ are constants, given by Eq.~(\ref{eqn:cjk_djk}). Hence $J_{jk}^{zz}$ can be tuned on-the-fly by controlling the relative strength of the two tones that determines $F_1/F_0$, as well as by changing the signs of $\delta_0,\delta_1$ (which are constrained to be equal in magnitude).
\begin{figure}[!htb]
    \centering
    \includegraphics[width=\columnwidth]{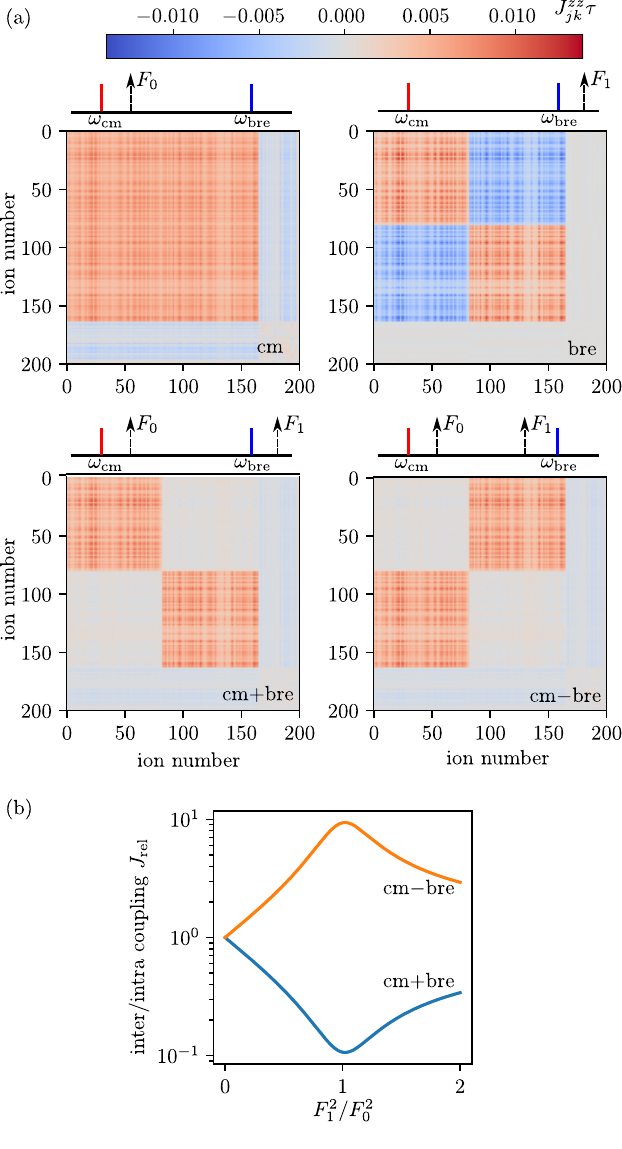}
    \caption{\textbf{On-the-fly control of intralayer and interlayer couplings using two-tone ODF.} (a) Matrix of ion-ion couplings $J_{jk}^{zz}\tau$ generated by a single tone coupling only to the center-of-mass mode (`cm'), single tone coupling only to the breathing mode (`bre'), two tones $F_0, F_1$ of equal strength ($F_0=F_1$) respectively coupling to each mode with identical detunings (`cm+bre') and with opposite detunings (`cm-bre'). The ion numbers are assigned based on whether an ion belongs to the top layer, bottom layer or the `scaffolding' structure, which respectively correspond to the ranges $0-81$, $82-163$ and $164-199$.  (b) Relative strength of interlayer to intralayer coupling, quantified by the metric $J_\mathrm{rel}$ [Eq.~(\ref{eqn:jrel})], versus the relative weights of the two tones $F_1^2/F_0^2$, for identical (`cm+bre') and opposite (`cm-bre') detunings. Here,  we use the crystal shown in Fig.~\ref{fig:anharmonic_eq}, ODF magnitude $F_0\approx 3.39\times10^{-23}$ N,  detunings $\abs{\delta_0}=\abs{\delta_1}\approx2\pi\times1.226$ kHz and $\tau=2\pi/\abs{\delta_0}\approx 816\;\mu$s.  
    }
    \label{fig:inter_intra}
\end{figure}

In Fig.~\ref{fig:inter_intra}(a), we illustrate some examples of the coupling matrices $J_{jk}^{zz}$ (multiplied by the evolution time $\tau$) realizable using the two-tone ODF described above. We tune $\theta_\mathrm{odf}$ so that $\Phi=0$, and first consider the case when $\delta_0,\delta_1>0$. When $F_0\ne 0,F_1=0$, only the c.m. contributes to the spin-spin interaction as in Sec.~\ref{sec:tunable_ising}, leading to positive coefficients between pairs of ions, irrespective of whether the ions belong to the same or different layers. A clear demarcation is visible in the matrix between the coupling coefficients for ion pairs belonging to the layer structure (square region approximately spanning ion numbers $0-160$ on each axis) and pairs where at least one ion belongs to the `scaffolding' structure of the crystal, see Fig.~\ref{fig:anharmonic_eq}. In contrast, when $F_0=0,F_1\ne 0$, only the breathing mode mediates the spin-spin interaction and its mode structure (see Fig.~\ref{fig:anharm_modes}) is clearly visible in the resulting coupling matrix: The intralayer couplings, which are on the block diagonal, are positive (orange) since all ions in a single layer move in phase. On the other hand, the interlayer couplings, which are on the block off-diagonal are negative (blue) since ions in different layers move out of phase. An advantage of using the breathing mode is that the scaffolding ions have negligible participation in this mode, leading to nearly vanishing couplings between the layer ions and the scaffolding ions.

The different form of the coupling matrices generated by the c.m. and breathing modes can be further exploited by simultaneously applying $F_0$ and $F_1$ as shown in the lower two panels of Fig.~\ref{fig:inter_intra}(a). When $F_0=F_1$ and $\delta_0,\delta_1>0$, the interlayer couplings induced by the c.m. and breathing modes strongly cancel each other. As a result, the coupling matrix of the bilayer crystal resembles that of two single layer crystals that do not interact with each other. In contrast, if $\delta_0>0$ and $\delta_1<0$, the intralayer couplings cancel, resulting in a bilayer with almost exclusively interlayer interactions. 

More generally, the relative strength of interlayer to intralayer coupling can be continuously tuned by controlling the ratio $F_1/F_0$ and the sign of $\delta_1$. As a measure of this relative strength, we define 
\begin{align}
    J_\mathrm{rel} = \frac{2\abs{\abs{J_{ud}}}_F}{\abs{\abs{J_{uu}}}_F+\abs{\abs{J_{dd}}}_F}.
    \label{eqn:jrel}
\end{align}
Here, the notation $\abs{\abs{M}}_F$ denotes the Frobenius norm of a matrix $M$, defined as $\abs{\abs{M}}_F=\sqrt{\sum_{jk}\abs{M_{jk}}^2}$. The matrices $J_{\alpha,\beta}$ appearing in Eq.~(\ref{eqn:jrel}), are sub-matrices of the couplings $J_{jk}$ between ion pairs where the first (second) ion belongs to layer $\alpha$ ($\beta$), with $\alpha,\beta\in\{u,d\}$ and $u,d$ denoting the upper and lower layers. In Fig.~\ref{fig:inter_intra}(b), we plot $J_\mathrm{rel}$ versus $F_1^2/F_0^2$ for $\delta_1>0$ (blue) and $\delta_1<0$ (orange). The two curves together demonstrate that the relative interlayer to intralayer coupling strength can be tuned over two orders of magnitude using the two-tone ODF technique described here. 

\subsubsection{Potential Applications: Tunable Bilayer Ising Models}

The ability to entangle spatially separated ensembles is an important requirement for several quantum information processing tasks. The Ising interactions studied in Secs.~\ref{sec:tunable_ising_control} and~\ref{sec:dynamic} can be used for the preparation of spin squeezed states relevant for quantum metrology applications~\cite{kitagawa1993PRA,ma2011PhysRep,pezze2018RMP}. The techniques described here can be used to prepare a variety of spin squeezed states with varying degree of intralayer and interlayer entanglement. These can range from a pair of approximately decoupled spin squeezed states in each layer, to the preparation of a global spin squeezed state involving ions in both layers. Such states may be relevant for distributed quantum metrology~\cite{komar2014NatPhys,guo2020NatPhys} as well as for the realization and characterization of two-mode squeezing, EPR (Einstein Podolsky-Rosen) correlations and EPR steering~\cite{duan2000PRL,julsgaard2001Nat,kurkjian2013PRA,byrnes2013PRA,fadel2018Sci}. Moreover, the capability to use large numbers of ions  also enables the use of the collective spin in each layer as a resource of long-lived and controllable continuous variables, as previously done with atomic vapors~\cite{Kuzmich2003}. This potentially opens a path to store and retrieve quantum information transmitted by phonons between different layers of ion arrays, as well as new avenues in quantum simulation. All of these applications are further facilitated by the ability to perform spin rotations and read out the spin projection in a layer-selective manner, which enables, for example, the detection and verification of bipartite entanglement~\cite{Vitagliano2023Quantum}.

Furthermore, the ability to dynamically tune the relative interlayer to intralayer coupling strength may enable the realization of a diverse gate set for variational quantum circuits~\cite{kokail2019Nat,kaubruegger2019PRL,kaubruegger2021PRX,marciniak2022Nat}. For instance, it was recently shown that the ability to implement both intra-ensemble and global entangling operations in a two-ensemble system is highly desirable for variational multiparameter quantum metrology~\cite{kaubruegger2023PRXQ}. The bilayer system may enable the study of such variational quantum metrology protocols on a large system consisting of hundreds of spins, where classical optimization is challenging.  
Additionally, applying a $\pi$ pulse only in one of the layers can allow for the change of  the sign of the interlayer  interactions for the performance of  time-reversal protocols that are relevant for the study of scrambling of quantum information~\cite{swingle2016PRA} and simulation of analogs of quantum gravity~\cite{Swingle2018}. For instance, the ability to realize similar intralayer couplings in both layers, and simultaneously vary the interlayer interaction may be useful for quantum simulation of worm-hole teleportation and operator spreading in controllable trapped ion arrays~\cite{Brown2023,Schuster2022}.

\subsection{Spin Exchange Models}

\subsubsection{Chiral Spin Exchange}
\label{sec:chiral}

The presence of a transverse field driving the spins can modify the Ising interaction induced by the ODF into a flip-flop interaction where pairs of spins exchange their excitations. As we discuss below, in bilayer crystals, the spin exchange can be made \emph{asymmetric}, i.e. the exchange coefficients produced by a uniform spin-dependent force along the $z$ direction [Fig.~\ref{fig:setup}(a)] can be complex, and hence the transfer of excitation from ion $j$ to $k$ and the reverse transfer from $k$ to $j$ can occur with opposite phases. Due to the broken directional symmetry, we refer to this process as a chiral spin exchange interaction. 

We consider the addition of a uniform transverse field with Rabi frequency $B_0$ that resonantly drives the $\ket{\downarrow}\leftrightarrow\ket{\uparrow}$ transition of the ions. This can be implemented straightforwardly with global microwave addressing of the crystal. In an interaction picture taken with respect to the free Hamiltonian of the spins and the normal modes, the interaction Hamiltonian has the form 
\begin{align}\label{eqn:ham_transverse_odf}
    \hat{H}_\mathrm{int.} = \sum_{j=1}^N \frac{\hbar B_0}{2}\hSig_j^x + 
     \sum_{j=1}^N F_0 \cos(\mu_r t+\phi_j) \hat{z}_j(t) \hSig_j^z.
\end{align}
To see the emergence of a spin-exchange model, it is useful to work in a spin space rotated about the $y$ axis. The spin operators in this rotated space are given by
\begin{align}
    \hat{\tau}_j^z = -\hSig_j^x,\; \hat{\tau}_j^x = \hSig_j^z. 
\end{align}
Assuming that the ODF difference frequency $\mu_r$ is tuned sufficiently far from any of the normal modes and provided that the transverse field $B_0$ is sufficiently strong (see Appendix~\ref{app:ChiralModels}), the normal modes can be adiabatically eliminated to arrive at an effective spin exchange model (denoted `ff' for flip-flop) of the form 
\begin{align}
    \hat{H}_\mathrm{eff}^\mathrm{ff} = \sum_{j=1}^N \frac{H_j}{2}\hat{\tau}_j^z + \sum_{\substack{j,k=1,j\neq k}}^N J_{jk}^\mathrm{ff}\hat{\tau}_j^+\hat{\tau}_k^-.
    \label{eqn:ham_ff}
\end{align}
The derivation of $\hat{H}_\mathrm{eff}^\mathrm{ff}$ is described in Appendix~\ref{app:ChiralModels}. Below, we focus on the form of the coefficients, especially, $J_{jk}^\mathrm{ff}$.

In order to discuss the coefficients, it is useful to introduce a normalized cross-amplitude $\mathcal{U}_{n,jk}\geq0$ and a relative motional phase $\varphi_{n,jk}$ for every pair of ions $j,k$, that characterize their motion along $z$ due to the $n$th normal mode. These quantities are defined by the equation  
\begin{align}
    \mathcal{U}_{n,jk}e^{i\varphi_{n,jk}} = \frac{u_{nj}^{z}(u_{nk}^{z})^*}{\dip{u_n^r}{u_n^r}}.
    \label{eqn:u_njk}
\end{align}
In addition, we introduce the detuning $\delta_n=\mu_r-\omega_n$ of the ODF difference frequency $\mu_r$ from the frequency of the $n$th normal mode. The coefficients for the $j=k$ terms appearing in Eq.~(\ref{eqn:ham_ff}) can then be expressed as 
\begin{align}
    H_j = \sum_n \frac{F_0^2 l_{0n}^2\mathcal{U}_{n,jj}B_0}{2\hbar^2(\delta_n^2-B_0^2)},
    \label{eqn:hj}
\end{align}
where $l_{0n}$ is the RMS zero-point displacement of the $n$th normal mode given by Eq.~(\ref{eqn:l0}). This expression assumes that the normal modes are initially in their quantum mechanical ground state, and an additional factor of $(2\nbar_n+1)$ appears for each mode when it is instead in a thermal state with mean occupation $\nbar_n$ (see Appendix~\ref{app:ChiralModels}). On the other hand, the $J_{jk}^\mathrm{ff}$ terms are independent of temperature, and have an expression of the form 
\begin{align}
    J_{jk}^\mathrm{ff} = J_{jk,r}^\mathrm{ff} + i J_{jk,i}^\mathrm{ff}
    \label{eqn:jkj_ff_1}
\end{align}
where the real and imaginary parts are given by 
\begin{align}
    J_{jk,r}^\mathrm{ff} &= \sum_{n} \frac{F_0^2 l_{0n}^2\mathcal{U}_{n,jk}\delta_n}{2\hbar^2(\delta_n^2-B_0^2)}\cos(\mathcal{D}_{n,jk}), \nonumber\\ 
    J_{jk,i}^\mathrm{ff} &= -\sum_{n} \frac{F_0^2 l_{0n}^2\mathcal{U}_{n,jk}B_0}{2\hbar^2(\delta_n^2-B_0^2)}\sin(\mathcal{D}_{n,jk}).
    \label{eqn:jkj_ff_2}
\end{align}
Here, we have introduced a phase $\mathcal{D}_{n,jk}$ defined as 
\begin{align}
    \mathcal{D}_{n,jk} = \Delta\phi_{kj}-\varphi_{n,jk},
    \label{eqn:d_njk}
\end{align}
where $\Delta\phi_{kj}$ is the ODF phase difference between ions $k$ and $j$, as described in Eq.~(\ref{eqn:phase_diff}).

Equation~(\ref{eqn:jkj_ff_1}) shows that the spin exchange is not symmetric, i.e. $J_{jk}^\mathrm{ff}\neq J_{kj}^\mathrm{ff}$, whenever $J_{jk,i}^\mathrm{ff}$ is non-zero. From Eq.~(\ref{eqn:jkj_ff_2}), a non-vanishing  $J_{jk,i}^\mathrm{ff}$ for an ion pair $j,k$ requires that $B_0\neq 0$, which is in any case a requirement for deriving $\hat{H}_\mathrm{eff}^\mathrm{ff}$, and additionally requires $\mathcal{D}_{n,jk}\neq 0,\pi$ for at least one mode $n$. As seen from Eq.~(\ref{eqn:d_njk}), the latter requirement can be satisfied in two ways: By coupling to a mode with a complex eigenvector such as, e.g., the mode $385$ shown in Fig.~\ref{fig:anharm_modes}, for which $\varphi_{n,jk}\neq 0,\pi$, or by controlling the ODF angle $\theta_\mathrm{odf}$ so that the phase difference $\Phi$ between ions $j,k$ in different layers of the bilayer crystals is different from $0$ or $\pi$.

Here, we will take the second approach described above and demonstrate how tuning $\theta_\mathrm{odf}$ can be used to engineer complex $J_{jk}^\mathrm{ff}$ coefficients. We assume that $\mu_r$ is tuned close to the highest frequency drumhead mode, which is the breathing mode, so that the contribution of other modes can be neglected. This is a reasonable assumption because we can arrange for $\delta_\mathrm{bre}=\mu_r-\omega_\mathrm{bre}$ and $B_0$ to be a few kilohertz while the next mode occurs around $50$ kHz below the breathing mode. Under these conditions, the intralayer spin-exchange coefficients are approximately real and positive, whereas the interlayer spin exchange coefficients are approximately of the form 
\begin{align}
    J_{jk}^\mathrm{ff} \propto  -\delta_\mathrm{bre} \cos(\Phi) + i B_0\sin(\Phi),
\end{align}
for ions $j$ and $k$ respectively belonging to the upper and lower layers. This expression has opposite signs compared to Eq.~(\ref{eqn:jkj_ff_2}) because of the breathing mode's interlayer motional phase $\varphi_{\mathrm{bre},jk}=\pi$.

\begin{figure}[!htb]
    \centering
    \includegraphics[width=\columnwidth]{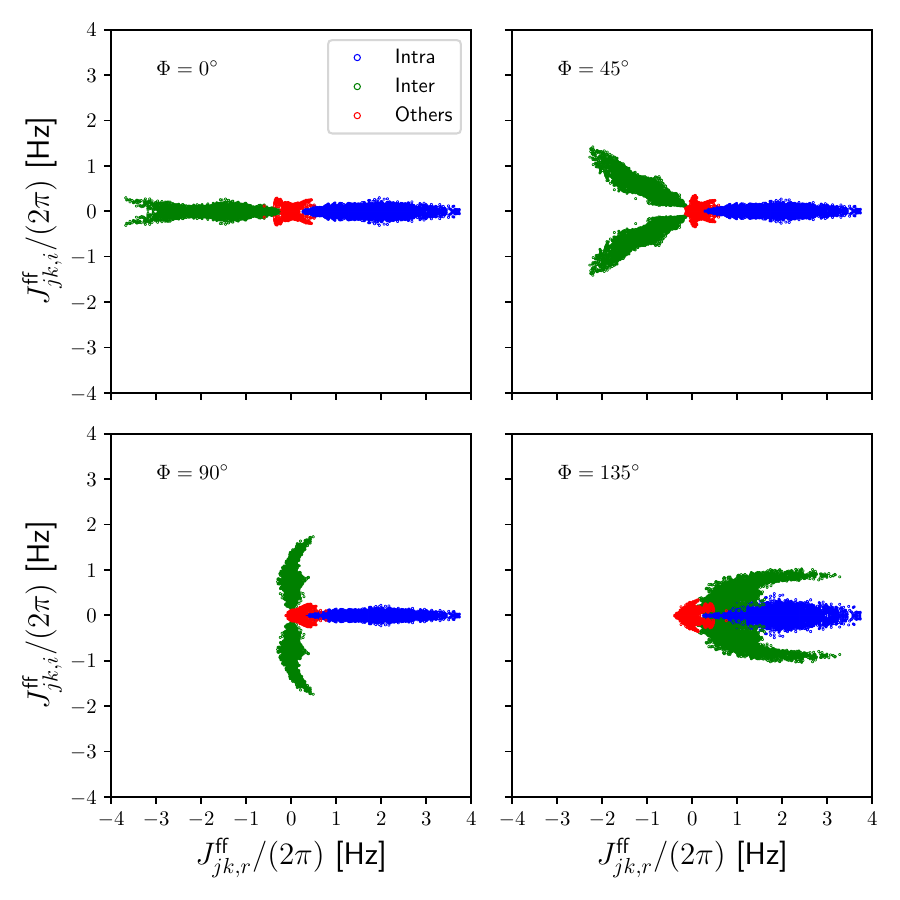}
    \caption{\textbf{Chiral spin exchange models.} Scatter plot of the spin-exchange coefficients $J_{jk}^\mathrm{ff}$ [Eq.~(\ref{eqn:jkj_ff_1})] in the complex plane, for all pairs of ions $j,k$, when the ODF difference frequency is tuned close to the breathing mode and an additional transverse field is present. For each pair, the markers are color coded according to whether the two ions belong to the same layer, different layers or have at least one ion that is not part of either layer.  The four panels show the variety of spin-exchange couplings possible for different interlayer phase differences $\Phi$, which can respectively be realized by setting $\theta_\mathrm{odf}\approx 0.81^\circ, 0.71^\circ, 0.61^\circ, 1.32^\circ$, resulting in $\Phi = 0^\circ, 45^\circ, 90^\circ, 135^\circ$. Here, we use 
    the crystal shown in Fig.~\ref{fig:anharmonic_eq}, ODF magnitude $F_0=6\times10^{-23}$ N,  detuning $\delta_\mathrm{bre}/(2\pi)=8$ kHz and transverse field strength $B_0/(2\pi)=4$ kHz. We note that, for these parameters, the ratio $(\delta_\mathrm{bre}-B_0)/(NJ)\approx 5.4$, where $J$ is a typical value of the coupling, and hence the system is not deep within the validity regime for adiabatic elimination of the breathing mode (see Appendix~\ref{app:ChiralModels}), which will therefore be excited to some extent in an experiment.     
    }
    \label{fig:jij_scatter}
\end{figure}

Figure~\ref{fig:jij_scatter} shows some of the possible structures of the spin exchange coupling coefficients as $\Phi$ is changed by controlling $\theta_\mathrm{odf}$. For each $\Phi$, the couplings between all ion pairs are shown as a scatter plot in the $J_{jk,r}^\mathrm{ff}-J_{jk,i}^\mathrm{ff}$ complex plane. For each pair, the markers are color coded according to whether the ions $j,k$ belong to the same layer (`intra'), different layers (`inter'), or have at least one ion from the scaffolding structure (`others'). As mentioned previously, the breathing mode does not have significant support in the scaffolding ions, and hence couplings to ions outside the two layers are small, as seen in all the panels of Fig.~\ref{fig:jij_scatter}. The intralayer couplings are predominantly real since all ions in a single layer experience approximately the same ODF phase; the small imaginary components arise because of the deviations of each layer from being a perfect plane. On the other hand, the interlayer couplings can be continuously tuned all the way from being predominantly real ($\Phi=0^\circ$) to almost purely imaginary ($\Phi=90^\circ$), highlighting the variety of chiral interlayer couplings possible in bilayer crystals. We note that although the values of the individual pairwise couplings shown here are rather small ($J_{jk}^\mathrm{ff}\sim 2\pi\times 4$ Hz), the interaction is all-to-all, and hence the collective dynamics can occur at appreciable rates ($NJ_{jk}^\mathrm{ff}\sim 2\pi \times 800$ Hz). 

\subsubsection{Potential Applications: Spin Exchange Models}

Spin exchange interactions emerging in the presence of a transverse field can potentially open several further directions for the applications of bilayer crystals. For instance, the anharmonicity of the trapping potential results in a c.m. mode that is not spatially homogeneous, leading to non-uniform spin-spin interactions across the crystal.  The inhomogeneity opens the possibility to enjoy full connectivity but in a model that breaks the permutation symmetry of the spins, thus allowing the system to explore the full Hilbert space without the speed bounds intrinsic to short range models. This could be an interesting avenue for the generation of fast quantum information scrambling~\cite{Belyansky2020}.  On the other hand, the
emergence of an energy gap from the collective interactions can impose energy penalties between the various
sectors, enabling studies of Hilbert space fragmentation~\cite{Khemani2020} under appropriate initial conditions.

Furthermore, the different layers constitute an additional degree of freedom that could  be used for the simulation of iconic models of orbital magnetism in solid state materials  possessing motional, orbital and charge degrees of freedom~\cite{Tokura2000}. The Jordan-Wigner transformation can be used to map the spin operators in each layer onto fermionic creation and annihilation operators describing interacting electrons that hop on a momentum-space lattice. The orbital degree of freedom, which arises in solids from the shape of the electron cloud, could be encoded in the layer degree of freedom.  The  implementation of orbital levels in ion crystals could provide  crucial insights on strongly correlated physics relevant for understanding a variety of phenomena in solids, such as metal-insulator transitions, high-temperature superconductivity and colossal magnetoresistance~\cite{Tokura2000}.

The charge degree of freedom that makes electrons respond to applied magnetic fields could be engineered by taking advantage of the chiral (complex) interlayer spin-exchange coefficients. The latter can be visualized as a complex hopping amplitude between the layers  via the  so called  Peierls substitution~\cite{Peierls1933}.  The chiral exchange emulates an effective magnetic field, or 
 more precisely,  an  analog of the Aharonov-Bohm phase which is proportional to the enclosed magnetic flux when a  particle circulates around a closed loop between the layers. Effective magnetic fields and complex-valued hopping amplitudes have 
been implemented on ultracold atom-based platforms  by using laser-assisted tunneling in an optical superlattice~\cite{Aidelsburger2011}, Raman and optical transitions~\cite{Kolkowitz2017,Bromley2018,Kolovsky2011,Mancini2015},  Floquet driving~\cite{Struck2012,Jotzu2014} and dipolar exchange interactions between Rydberg atoms~\cite{Lienhard2020,chen2023strongly}. Alternative platforms have also been realized using superconducting qubits~\cite{Roushan2017}  and photonic~\cite{Ozawa2019}  or phononic~\cite{Liu2020} systems.  The implementation of a synthetic gauge field in ion crystals will open exquisite opportunities to study new forms of topological states of matter under controllable conditions.

Alternatively, in the context of spintronics, the complex interlayer exchange coefficients demonstrated in Sec.~\ref{sec:chiral} may enable the simulation of models with antisymmetric Dzyaloshinskii–Moriya DM interactions~\cite{vedmedenko2019PRL}. To see this, we note that the Hamiltonian~(\ref{eqn:ham_ff}) can be written as 
\begin{align}
    \hat{H}_\mathrm{eff}^\mathrm{ff} = &\sum_{j=1}^N \frac{H_j}{2}\hat{\tau}_j^z + \sum_{j, k=1,k<j}^N\frac{J_{jk,r}^\mathrm{ff}}{2}\left(\hat{\tau}_j^x\hat{\tau}_k^x+ \hat{\tau}_j^y\hat{\tau}_k^y \right) \nonumber\\
    & +\sum_{j,k=1,k<j}^N\frac{J_{jk,i}^\mathrm{ff}}{2}\left(\hat{\tau}_j^x\hat{\tau}_k^y- \hat{\tau}_j^y\hat{\tau}_k^x \right).
\end{align}
While the second term in the first line correspond to an XY model, the second line corresponds to an antisymmetric DM interaction $\mathbf{D}_{jk}\cdot(\hat{\mbf{\tau}}_j \times \hat{\mbf{\tau}}_k)$ with a DM vector $\mathbf{D}_{jk}$ along the $z$ axis in spin space.

\section{Practical considerations}
\label{sec:practical}

In this section, we briefly analyze a number of factors that can potentially limit the fidelity of quantum protocols using bilayer crystals. We also outline possible strategies to mitigate adverse affects.

\subsection{Off-Resonant Light Scattering}

The spin-dependent force $F_0$ is realized using lasers that couple the spin states of each ion to higher electronic states via dipole-allowed transitions~\cite{sawyer2012PRL}. Consequently, the same lasers also induce scattering of photons into free space, that can preserve the spin state (Rayleigh scattering) or flip it (Raman scattering)~\cite{carter2023PRA}. This scattering rate $\Gamma$ is proportional to the laser power $P$ used to generate $F_0$. For a fixed $F_0$, the  required laser power $P$ is approximately proportional to $1/\theta_\mathrm{odf}$. In this work, we assumed a small ODF angle $\theta_\mathrm{odf}\sim 1^\circ$ in order to ensure a nearly uniform ODF phase across a single layer of the bilayer crystal. Typical applications use a larger incidence angle $\theta_\mathrm{odf}\sim 10^\circ$. Consequently, the expected decoherence due to off-resonant light scattering at $\theta_\mathrm{odf}\sim 1^\circ$ is expected to be $10\times$ larger than under typical operating conditions. 

Nevertheless, a number of strategies can be used to decrease $\Gamma$ for a fixed $F_0$. One possible route is to parametrically amplify the spin-dependent force, which will allow for the use of lower laser powers for the same $F_0$~\cite{ge2019PRL,burd2021NatPhys}. We note that, although this technique has been demonstrated for 1D chains and 2D crystals of ions, its generalization to a bilayer or a general 3D crystal requires further careful considerations, which we leave to future work.  A second option is to simply operate at a higher $\theta_\mathrm{odf}$, which reduces the required $P$ at the cost of reducing the bilayer quality. Moreover, drawing parallels to previous studies on the areal density of single-plane crystals~\cite{Dubin2013PRA}, it is possible that the inclusion of higher-order anharmonic terms in the trapping potential, such as a $C_6$ term, may further reduce the thickness of each individual layer and hence allow for larger $\theta_\mathrm{odf}$ without compromising on the bilayer quality. We note, however, that the optimal $C_6$ term magnitude may be large; nevertheless, the trap electrode structure can be designed to boost the anharmonic terms. Finally, we note that decoherence due to spontaneous emission can be mitigated by using high laser intensities that enable the implementation of spin-dependent forces with large detunings from resonant transitions~\cite{carter2023PRA}.  

We note that even in the worst-case scenario where decoherence precludes operation at small $\theta_\mathrm{odf}$, quantum protocols with bilayer crystals at $\theta_\mathrm{odf}\sim10^\circ$ may still be of great value. Although the ODF lattice no longer `resolves' the individual layers, i.e., the ODF phase is not homogeneous across each layer [see Fig.~\ref{fig:setup}(b)], they are nevertheless resolvable as two distinct layers in the side-view readout, and hence the capability to detect and characterize bipartite correlations and entanglement in spatially separated ensembles is unaffected. Furthermore, it may be possible to realize effective collective interactions even if the underlying interactions are inhomogeneous, by taking advantage of gap protection ideas or dynamical decoupling protocols~\cite{perlin2020PRL,franke2023Nat}. 

\subsection{Thermal Motion}

Residual thermal motion of ions can impact quantum protocols in trapped ion systems in multiple ways. Here, we focus on three effects that may be particularly important for bilayer crystals. 

\subsubsection{Lamb-Dicke Confinement}

Spin-motion coupling protocols require the RMS amplitude of thermal motion of the ions along the ODF lattice direction ($z$ axis) to be small compared to the lattice wavelength $2\pi/\Delta k$. The quantity to assess the confinement is the parameter $\eta_j = \Delta k \sqrt{\ev{\hat{z}_j^2}}$, for which we require $\eta_j\ll 1$~\cite{wineland1998JResNIST}. For a crystal in thermal equilibrium at temperature $T$, 
\begin{align}
    \ev{\hat{z}_j^2} = \sum_{n=1}^{3N} l_{0,n}^2\frac{\abs{u_{nj}^{z}}^2}{\dip{u_n^r}{u_n^r}}(2\nbar_n(T)+1),
    \label{eqn:zrms_thermal}
\end{align}
where $\nbar_n(T)=[\exp(\hbar\omega_n/(k_B T))-1]^{-1}$ is the thermal occupation of mode $n$. For the same $T$, the typical $\eta_j$ values are larger for a bilayer than for a single-plane crystal. This increase can be attributed primarily to the $\exb$ modes, which have high thermal occupation on account of their low frequencies, and acquire a non-zero mode fraction along the $z$ direction in the bilayer case, as shown in Fig.~\ref{fig:w_R_fz}. Nevertheless, a small ODF angle of $\theta_\mathrm{odf}\sim 1^\circ$ ensures that $\eta_j\lesssim 0.1$ for $T\approx450\; \mu$K ($\nbar_{\mathrm{cm}}\approx 5$), which corresponds to the Doppler cooling limit of the drumhead modes, and $\eta_j\lesssim 0.035$ for $T\approx 50\;\mu$K ($\nbar_{\mathrm{cm}}\approx 0.3$), which corresponds to the temperature after near ground-state cooling of the drumhead modes~\cite{Jordan2019PRL}. We note that these estimates assume parameters relevant for recent NIST experiments and that \emph{all} the modes are in thermal equilibrium at the specified $T$, although these temperatures have only been measured on the drumhead modes. In single plane crystals, the $\exb$ modes are typically poorly cooled and at much higher temperatures of the order of $10$ mK. However, we discuss prospects for their near ground-state cooling in Sec.~\ref{sec:cooling_prospects}.

\subsubsection{Mode Frequency Fluctuations}

Thermal motion associated with low-frequency $\exb$ modes leads to large-amplitude position fluctuations of the ions about their equilibrium positions. In turn, these fluctuations lead to spectral broadening of the drumhead modes~\cite{shankar2020PRA}, which reduces the fidelity of quantum protocols. In a purely harmonic trapping potential, the c.m. mode is an exception as its frequency is insensitive to position fluctuations. This is because the ion displacements due to this mode are all identical, which decouples this mode from the anharmonic effects arising from the Coulomb interaction. However, the preparation of clean bilayers requires an explicit anharmonic component in the trapping potential, in which case the ion displacements due to the c.m. mode are no longer uniform across the crystal. Hence, the c.m. mode frequency is also sensitive to ion position fluctuations in anharmonic traps. To estimate the extent of spectral broadening, we use a procedure similar to the `thermal snapshot' analysis carried out in Ref.~\cite{shankar2020PRA}. For the crystal and trap parameters chosen in Fig.~\ref{fig:anharmonic_eq}, and assuming $T\approx 50\;\mu$K, we estimate that the frequency fluctuations in the c.m. and breathing modes can respectively be of the order of $200$ Hz and $1$ kHz respectively. The primary contributor appears to be the lowest frequency $\exb$ mode, which is a rocking mode with frequency $262$ Hz and $\nbar_n\approx 4000$ at $T\approx 50\;\mu$K. The frequency of this mode can be increased by using a stronger rotating wall, i.e. by increasing $\delta$. Using $\delta\approx 0.0126$, we find that the frequency fluctuations of the c.m. and breathing modes are reduced to $10$ Hz and $160$ Hz respectively. Further cooling of the $\exb$ modes to lower their thermal occupation to $\nbar_n\sim \mathcal{O}(1)$ can bring these frequency fluctuations to the $1$ Hz level. A detailed discussion of this analysis and its extensions to general 3D crystals with large numbers of ions will be presented in a future work. 

\subsubsection{Higher order spin-motion coupling}

The spin-motion coupling~(\ref{eqn:hodf_1}) induced by the ODF is approximately linear in the ion displacement [Eq.~(\ref{eqn:hodf_2})] only if the ions are confined sufficiently deep in the Lamb-Dicke regime, i.e., $\Delta k \ev{\hat{z}_j^2}^{1/2}\ll 1$, where $\ev{\hat{z}_j^2}$ is a function of temperature as given in Eq.~(\ref{eqn:zrms_thermal}). The adverse affects of spin-motion coupling terms of order $\hat{z}_j^2$ or higher may become important when the modes are at a non-zero temperature. These terms lead to residual spin-motion entanglement at decoupling times and also modify the effective spin-spin coupling coefficients, and can be particularly detrimental if the sum or difference frequencies of two normal modes accidentally land on or near resonance with the ODF difference frequency $\mu_r$. Compared to typical single-plane crystals, accidental resonances may be more pronounced in bilayer crystals because of two reasons. First, the finite mode fraction of the $\exb$ modes along the $z$ direction can lead to accidental resonances because of drumhead-$\exb$ mixing. Second, the increased bandwidth of the drumhead modes can result in the sum frequency of two low-frequency drumhead modes falling close to $\mu_r$, which is typically tuned close to one of the high-frequency drumhead modes.  Our preliminary estimates suggest that for $\theta_\mathrm{odf}\sim 1^\circ$, and for a temperature $T\lesssim 50\;\mu$K, the corrections to the effective spin-motion and spin-spin coupling coefficients from the $\hat{z}_j^2$ terms are small. However, a critical assessment of the detrimental effects of higher-order terms in the ODF interaction, and identifying conditions under which they can be suppressed, is ultimately a task for experiments, given the vast number of normal modes intrinsic to large trapped ion crystals.

\subsubsection{Prospects for Near Ground-State Cooling}\label{sec:cooling_prospects}

The above practical considerations suggest that near ground-state cooling of all the modes of the crystal may be critical for implementing high-fidelity quantum protocols with bilayer crystals. Efficient Doppler cooling, and even near ground-state cooling of the drumhead modes using electromagnetically induced transparency (EIT), have already been demonstrated for single-plane crystals~\cite{Jordan2019PRL}. Theory and numerical work can explore whether these techniques are readily applicable to bilayer and general 3D crystals. The high-frequency cyclotron modes are efficiently Doppler cooled to occupations of a few quanta~\cite{torrisi2016PRA,tang2019POP}, and moreover, their high frequency and small orbits make them effectively decoupled from the other mode branches~\cite{shankar2020PRA,tang2021PRA}. The main challenge is to design efficient Doppler and sub-Doppler cooling schemes for the low-frequency $\exb$ modes. In single-plane crystals, Doppler and sub-Doppler cooling of the $\exb$ modes is complicated by the crystal rotation, which makes access to the planar normal modes challenging. Recent numerical studies have shown that it is possible to sympathetically cool the $\exb$ modes of single-plane crystals to around $1$ mK by resonantly coupling them with low frequency drumhead modes~\cite{johnson2023arXiv}. Although such a technique may not directly apply to a bilayer crystal, where the two branches have a frequency gap, a resonant coupling can still be achieved by the application of suitable `axialization' potentials~\cite{powell2002PRL,hendricks2008JPhysB}.  However, an advantage of bilayers that is absent in single-plane crystals is the non-zero component of the $\exb$ modes along the $z$ direction, which opens a route to address them without coupling to the crystal rotation. As a result, it may be possible to use EIT or other sideband cooling techniques to directly cool the $\exb$ modes close to their motional ground state. We note that sideband cooling has previously been demonstrated on the $\exb$ mode of a single ion in a Penning trap~\cite{hrmo2019PRA}. Another factor that may facilitate the cooling of these modes is that the $\exb$ modes rapidly equilibrate with each other~\cite{tang2021PRA}. As a result, efficient cooling of even a small part of the $\exb$ mode branch may lead to strong sympathetic cooling of the entire branch. Techniques alternative to laser cooling may also help to mitigate ion position fluctuations, such as the use of co-rotating optical tweezers for pinning certain ions and stabilizing the crystal structure~\cite{laupretre2019PRA}. 

\subsection{Frequency Resolution}\label{sec:FrequencyResolution}

For large ion crystals, frequency crowding of modes can occur because of the large number of modes and limited bandwidth. A large spectral isolation of the modes is desirable for a number of quantum protocols to ensure that residual spin-motion coupling from nearby spectator modes is small. For the crystal shown in Fig.~\ref{fig:anharmonic_eq}, the breathing mode is the highest frequency drumhead mode and it is well separated by approximately $50$ kHz from the next drumhead mode. The separation of the c.m. mode from its nearest mode is smaller, around $15$ kHz. However, for sufficiently small detunings from the c.m. mode, such as the value $\delta_0/(2\pi)=(\mu_r-\omega_\mathrm{cm})/(2\pi)\sim 1$ kHz used in Sec.~\ref{sec:tunable_ising}, the residual spin-motion coupling arising from nearby modes can be neglected in a first approximation. Moreover, the impact of frequency crowding can in principle be mitigated by frequency and amplitude modulation of the spin-dependent force. Such techniques have been shown to enable high-fidelity generation of target entangled spin states in the presence of crowded mode spectra in small crystals~\cite{choi2014PRL,leung2018PRL,shapira2020PRA}.

\section{Conclusion and Outlook}
\label{sec:conclusion}                                      

\begin{figure}[h]
    \centering
    \includegraphics[width=0.9\columnwidth]{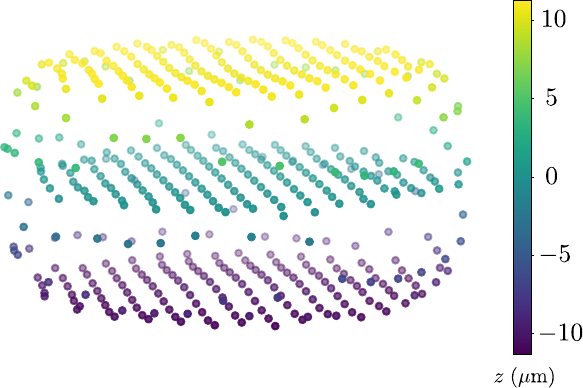}
    \caption{\textbf{Multilayer crystals beyond bilayers.} 3D view of a numerically obtained equilibrium configuration of a trilayer crystal with $N=500$ trapped ions in a Penning trap. Here, we set $\omega_r/(2\pi)=215$ kHz, $\cfb=1.659$, and take other parameters to be the same as in Fig.~\ref{fig:harmonic_eq}.
    }
    \label{fig:trilayer}
\end{figure}

The key contribution of our work is the demonstration of a path for quantum information processing with \emph{structured} 3D crystals of large numbers of trapped ions, which have hitherto been restricted to 1D chains or planar 2D crystals. Although our work has focused specifically on the realization and applications of a bilayer crystal, we note that it may be possible to extend the ideas presented here to realize multi-layered trapped ion crystals. As an illustration, Fig.~\ref{fig:trilayer} shows an equilibrium \emph{trilayer} crystal consisting of $N=500$ ions in the presence of an anharmonic trapping potential. The realization of multilayered crystals beyond bilayers opens further opportunities to use trapped ion systems to probe exotic 3D phenomena, such as the chiral transport of spin excitations across a multi-layer array of atoms~\cite{bilitewski2023PRL}.

The capabilities discussed in Sec.~\ref{sec:prospects} by no means constitute an exhaustive toolbox for quantum information processing with bilayer crystals. For instance, our examples here focused on the realization of various types of effective spin-spin couplings. By probing the system away from the decoupling times, or by coupling $\text{(near-)}$ resonantly with a normal mode, it is also possible to realize and explore a variety of spin-boson models~\cite{porras2004PRL,safavinaini2018PRL}. The bilayer configuration can greatly enrich the variety of spin-boson models that can be studied, by allowing for layer-resolved state preparation and readout, as well as the realization of chiral spin-motion coupling by manipulating the laser incidence angle or by coupling to chiral drumhead modes. Moreover, since the mode can be brought closer to resonance for simulating spin-boson models, the spin-motion coupling can be larger and can thus speed up the simulation, thereby mitigating the adverse impacts of off-resonant light scattering and higher-order terms in the ODF interaction, which were discussed in Sec.~\ref{sec:practical}. We also note that the layer-resolved addressing allows the study of dissipative protocols and open quantum systems~\cite{barreiro2011Nat,bermudez2013PRL,reiter2016PRL,shankar2017PRA}, e.g., by engineering dissipation in one layer and coupling it to the second layer, which undergoes purely coherent evolution. Furthermore, it is possible to introduce individual ion addressing in this system by exploiting the staggered positions of ions in the different layers to address individual ions using tightly focused laser beams along the $z$ axis, or by using optical setups based on deformable mirrors~\cite{polloreno2022PRR}. 

In this work, we explored the use of the light-shift gate to engineer an optical dipole force. An alternative strategy is to employ a M\o lmer-S\o rensen (MS) gate, which can be configured in two different ways that lead to different spin-motion couplings~\cite{lee2005, haljan2005}. This can enable us to, e.g., couple orthogonal quadratures of a single normal mode to the same component of the collective spin in each layer, or alternatively, couple a different component of the collective spin in each layer to the same quadrature of a single normal mode. Furthermore, along with an appropriate set of single qubit rotations, one of the MS gate configurations can lead to dynamics that is effectively independent of $\phi_j$, or equivalently, the ion positions. As a result, uniform entangling operations can be performed using this gate for \emph{any} ion crystal structure without loss of contrast, and they can also be performed at larger incidence angles in order to suppress spontaneous emission errors. The MS gate could therefore be an additional powerful tool for expanding the possible quantum operations in both bilayer and larger 3D crystals in Penning traps.

More broadly, 3D crystals in Penning traps enable us to use a well-established approach, viz., engineering spin-motion coupling using 1D optical lattices along the $z$ direction, to open up new opportunities in quantum information processing. At the same time, complementary efforts are required from a non-neutral plasma physics perspective to understand the limits for the preparation and control of large ion crystals for quantum protocols. For instance, our work motivates the need to rigorously understand the role of anharmonicity in the realization of clean multi-layered crystal configurations and their long-term stability which may be impacted by processes such as background gas collisions. Furthermore, given the adverse affects of thermal motion, our work also underscores the urgent need to design efficient Doppler and sub-Doppler cooling techniques for full 3D cooling of large ion crystals, which may be possible using molecular dynamics simulations. Finally, we hope that our work can inspire analogous research into preparing structured 3D crystals in rf Paul traps, where anharmonic trapping potentials have previously been employed to produce more uniformly spaced 1D chains of trapped ions~\cite{lin2009EPL}.

\section*{acknowledgments}

We thank M. Miskeen Khan, Jennifer Lilieholm, and Wes Johnson for a careful reading and feedback on the manuscript. We acknowledge discussions with Dan Dubin, John Zaris and Scott Parker. S.H. acknowledges the support of Kishore Vaigyanik Protsahan Yojana, Department of Science and Technology, Government of India.  A.S. acknowledges the support of a C.V. Raman Post-Doctoral Fellowship, IISc.
A.L.C., A.M.R. and J.J.B.  acknowledge funding from  the U.S. Department of Energy, Office of Science, NQI Science Research Centers, Quantum Systems Accelerator (QSA), a collaboration between the U.S. Department of Energy, Office of Science and other agencies.  A.M.R. acknowledges additional support from VBFF,  ARO grant  W911NF-16-1-0576,  by the NSF JILA-PFC PHY-2317149, QLCI-OMA-2016244, and by NIST. J.J.B. acknowledges additional support from the DARPA ONISQ program and AFOSR grant FA9550-201-0019.

\appendix

\section{Algorithm used for equilibriating the crystal}\label{app:Modified_Basin_hopping}

For an $N$ ion crystal, the equilibrium configuration corresponds to determining the $3N$ position coordinates of the crystal that minimizes the total potential energy given by $U=\sum_j e\phi_j$, where $e\phi_j$ is given by the sum of Eqs.~(\ref{eqn:harmonic_phi}) and (\ref{eqn:anharmonic_phi}).  In Ref.~\cite{wang2013PRA}, a local gradient descent optimization was employed to find the equilibrium configuration of single-plane crystals, starting from a seed lattice inspired from experimental observations. However, this approach is not directly applicable for large 3D crystals, where it is challenging to identify the best crystal configuration because of the sheer number of local minima and plateaus in which a solution can get trapped. Therefore, in this work, we employ a variation of a \emph{global} optimization technique called the Basin-Hopping algorithm.

A pseudocode for our minimization routine is described in Algorithm~\ref{alg:modified_basin_hopping}. Standard basin-hopping involves repeatedly (i) nudging the current local minimum $\mathbf{x}_\mathrm{old}$, (ii)  determining a new local minimum $\mathbf{x}_\mathrm{new}$ starting from the nudged solution using a gradient-based method, and (iii) choosing between the current and new local minima based on a fixed, preset temperature $T$ using a Metropolis-Hastings acceptance step. Here, we use the Newton-Conjugate Gradient (NCG) method as our local minimization routine. The standard basin-hopping method has been shown to be advantageous in landscapes where a single large valley is additionally modulated by several local peaks and valleys. In order to account for finer structure in these local peaks and valleys, we modify the standard basin-hopping algorithm to include a cooling (annealing) schedule on the temperature $T$, which controls both the nudge step size as well as the acceptance probability for the new local minimum. Our routine stops when the annealing temperature reaches $0$, and the solution $\mathbf{x}_\mathrm{best}$ with the lowest potential energy $U$ is returned as the equilibrium configuration.   

\begin{algorithm}[h]
\caption{Modified Basin-Hopping\\
For the minimization, we used $n_\text{steps} = 20$, $T_\text{start} = 0.048$, $\alpha_x = 1$ and the position vector $\vb{r}$ and energy $U$ are made dimensionless by performing the scaling $\vb{r} = \vb{x} \cdot l_0$, $U \to E_0 \td{U}$ where $ l_0 = \left(\frac{2k_e e^2}{m\omega_z^2}\right)^{1/3}$, $E_0 = \frac{1}{2} m l_0^2 \omega_z^2$. The notation $\text{NCG}(\td{U},\vb{x})$ refers to a Newton-Conjugate Gradient method applied using $\td{U}$ as the cost function to be minimized and $\vb{x}$ as the initial point.}\label{alg:modified_basin_hopping}
$\vb{x}_\text{new} \gets \text{Unif}^{\otimes 3N}(-\sqrt[3]{N}, \sqrt[3]{N})$\;
$\alpha \gets \text{Nudge Scale}$\;
$\vb{x}_\text{new} \gets \text{NCG}(\td{U},\vb{x}_\text{new})$\;
$\vb{x}_\text{best} \gets \vb{x}_\text{new}$
\For{$T \gets \text{linspace}(T_\text{start}, 0, n_\text{steps})$}{
  $\vb{x}_\text{old} \gets \vb{x}_\text{new}$\;
  $\vb{x}_\text{new} \gets \text{NCG}(\td{U},\vb{x}_\text{new} + \mathcal{N}^{\otimes 3N}(0, \alpha T))$\;
  \eIf{$\exp((\td{U}(\vb{x}_\text{old}) - \td{U}(\vb{x}_\text{new}))/T) > {\text{Unif}}(0,1)$}{
  \If{$\td{U}(\vb{x}_\text{new}) < \td{U}(\vb{x}_\text{best})$}{
    $\vb{x}_\text{best} \gets \vb{x}_\text{new}$\;
  }
  }{
  $\vb{x}_\text{new} \gets \vb{x}_\text{old}$\;
  }
}
\Return{$\vb{x}_\text{best}$}
\end{algorithm}

In order to account for the potential energy landscape, the nudge scale $\alpha$ is chosen to be different along the $x,y$ and $z$ directions, such that $\alpha_x = \alpha_y = \alpha_z/\beta$, where $\beta$ is given by Eq.~(\ref{eqn:beta}). The value of $\alpha_z$ is chosen such that for a large range of temperatures, the acceptance probability is on an average $0.5$. 

We tested our algorithm against a simpler approach where a new minimum is accepted only if it is lower in energy than the current minimum, which corresponds to standard basin-hopping at $T=0$. We observed that, for the same number of local minimization calls, both algorithms reach configurations with similar final energies, however, our algorithm leads to a $3\times$ reduction in the variance of the final energies attained. In other words, it is more consistent in finding final configurations with lower energy.

\section{Anharmonicity in Penning traps}\label{app:AnharmonicityExperimental}

Here, we study the dependence of the dimensionless coefficient $\cfb$ on the various trap parameters to understand how optimal $\cfb$ values can be achieved in practice. For cylindrical trap electrodes with perfect azimuthal symmetry, the potential arising from the electrodes can be expressed at a point $x,y,z$ in the trap in terms of Legendre polynomials $P_n$ as~\cite{Dubin2013PRA}
\begin{equation}
    V(r) = \sum_{n=0}^\infty V_{2n} (r/d)^{2n} P_{2n}(z/r),
    \label{eqn:multipole_exp}
\end{equation}
where $r = \sqrt{x^2 + y^2 + z^2}$ is the distance from the trap center (taken as the origin), $d$ is a characteristic trap length scale (typically the trap electrode radius), and the $V_{2n},n=1,2,\ldots$ are coefficients of the multipole expansion with units of voltage. Here, we assume that the odd terms in the voltage are heavily suppressed, and hence can be ignored. This can be ensured by appropriate trap design and applied voltages. The $V_{2n}$ coefficients can be related to the voltages applied at the trap electrodes through a linear calibration~\cite{gabrielse1983, beaty1986, fei1999}. The potential~(\ref{eqn:multipole_exp}) can be explicitly written up to fourth order as 
\begin{equation}
    V(\rho,z) = V_0 + \frac{V_2}{d^2} (z^2 - \frac{1}{2}\rho^2) + \frac{V_4}{d^4} (z^4 - 3z^2\rho^2 + \frac{3}{8}\rho^4),
    \label{eqn:multipole_exp_2}
\end{equation}
where $\rho = \sqrt{x^2 + y^2}$ is the distance from the trap axis.

Comparing Eqs.~(\ref{eqn:anharmonic_phi}) and~(\ref{eqn:multipole_exp_2}), we can relate 
\begin{align}
    \cfb 
    &\propto \frac{N^{2/3} V_4}{d^4 m^{5/3} \omega_z^{10/3} \beta^{5/3}}.
\end{align}
This suggests that to achieve a desired value of $\cfb$ keeping $\omega_z,\beta$ fixed, the required $V_4$ can be decreased by either increasing the number of ions, or by decreasing the trap size.  

Furthermore, since the bilayer regime is characterized not by the value of $\beta$, but by the relative value $\beta/\beta_c$, it is reasonable to fix $\beta/\beta_c$ instead of just $\beta$. Since $\beta_c\propto1/\sqrt{N}$, the scaling of $\cfb$ then becomes superlinear in $N$, i.e.,
\begin{align}
    \cfb &\propto \frac{N^{3/2}V_4 }{d^4 m^{5/3} \omega_z^{10/3} (\beta/\beta_c)^{5/3}},
\end{align}
suggesting that the required $V_4$ can be greatly reduced by working with larger ion crystals. For instance, for the same $\cfb$, a crystal with $N=1000$ ions requires a $11\times$ smaller $V_4$ compared to an $N=200$ ion crystal. Preliminary numerical simulations that we have performed indeed provide evidence for a superlinear reduction with $N$ of the $V_4$ values required to achieve clean bilayer crystals, although the numerically observed scaling exponent is somewhat smaller than $3/2$. 

\section{The Stiffness Matrix}\label{app:StiffnessMatrix}

The stiffness matrix $\K$ of the system is a $3N\times 3N$ real, symmetric 
matrix that can be written as 9 sub-matrices in the following form: 
\begin{align}
\K = \begin{pmatrix}
  \K^{xx} & \K^{xy} & \K^{xz} \\
  \K^{yx} & \K^{yy} & \K^{yz} \\
  \K^{zx} & \K^{zy} & \K^{zz} \\
\end{pmatrix},
\end{align}
where $\K^{\alpha\gamma}$ for $\alpha, \gamma \in \{x,y,z\}$ is an $N\times N$ matrix, such that $\K^{\alpha\gamma} = \K^{\gamma \alpha}$. Additionally, each sub-matrix is symmetric in the ion indices, i.e. $\K_{jk}^{\alpha\gamma} = \K_{kj}^{\alpha\gamma}$.

The element $\K_{jk}^{\alpha\gamma}$ can be determined from the Hessian of the potential energy as 
\begin{align}
    \K_{jk}^{\alpha\gamma} = \frac{1}{2}\pdv{U}{\alpha_j}{\gamma_k},
\end{align}
where $U=\sum_j e\phi_j$ is the total potential energy of the system, and $e\phi_j$ includes the harmonic and anharmonic terms given by Eq.~(\ref{eqn:harmonic_phi}) and Eq.~(\ref{eqn:anharmonic_phi}). The contribution to $\K$ from the Coulomb interaction leads to cross terms between ions, while the contribution to the external potential is only to terms of the form $\K_{jj}$. So, it is useful to rewrite $\K$ as
\begin{equation}
    \K^{\alpha\gamma}_{jk} = C^{\alpha\gamma}_{jk} (1-\delta_{jk}) + \left(E^{\alpha\gamma}_j - \sum_{i\ne k} C^{\alpha\gamma}_{ik} \right) \delta_{jk},
\end{equation}
where the $C$-matrix denotes the off-diagonal contribution due to the Coulomb interaction,  and $E$ denotes the contribution due to the external field.
The explicit expressions for the different elements are found to be as follows:
\begin{align}
    C^{\alpha\gamma}_{jk} &= \frac{k e^2}{2 r_{kj}^3} \left(\delta_{\alpha \gamma} - 3\frac{(\alpha_k - \alpha_j)(\gamma_k - \gamma_j)}{r_{kj}^2} \right) ,\\
    E^{xx}_{j} &= \frac{m\omega_z^2}{2}\left[(\beta + \delta) + \frac{\beta \cfb}{r_{p,0}^2}\left(- 6z_j^2 + \frac{9}{2}x_j^2 + \frac{3}{2} y_j^2\right)\right],\\
    E^{yy}_j &= \frac{m\omega_z^2}{2}\left[(\beta - \delta) + \frac{\beta \cfb}{r_{p,0}^2} \left(- 6z_j^2 + \frac{9}{2}y_j^2  + \frac{3}{2} x_j^2\right)\right],
\end{align}
\begin{align}
    E^{zz}_j &= \frac{m\omega_z^2}{2}\left[1 + \frac{\beta\cfb}{r_{p,0}^2}\left(12z_j^2 - 6\left(x_j^2 + y_j^2\right)\right)\right],\\
    E^{xy}_j &= -\frac{3}{2}m\omega_z^2\frac{\beta\cfb}{r_{p,0}^2} x_ky_k,\\
    E^{yz}_j &= -6m\omega_z^2\frac{\beta\cfb}{r_{p,0}^2}y_kz_k,\\
    E^{zx}_j &= -6m\omega_z^2\frac{\beta\cfb}{r_{p,0}^2}z_kx_k.
\end{align}

\section{Details of Quantization of Normal Modes}\label{app:ModeQuantization}

As described in Sec.~\ref{sec:Quantization}, we quantize the normal modes of the system by assuming an ansatz for the quantized modes, and showing that it leads to the correct canonical commutation relations for the ion positions and canonical momentum operators. In this appendix, we demonstrate this for the $x,p^x$ operators of the ions. A similar procedure can be used to verify the consistency of the approach for the $y,z$-coordinates. To quantize the system, we first find the explicit form of the canonical momenta to be given by
\begin{equation}
    p^x_j = \pdv{L}{\dot{x}_j} = m\dot{x}_j - m\frac{\omega_c'}{2}y_j,
\end{equation}
which can be written in terms of the complex amplitudes $A_n$ of the normal modes using Eq.~(\ref{eqn:q_vec}). Making the substitution
\begin{equation}
A_n \rightarrow \ha_n \frac{l_{0n}}{\sqrt{\dip{u_n^r}{u_n^r}}},\ \ A_n^{*} \rightarrow \had_n \frac{l_{0n}}{\sqrt{\dip{u_n^r}{u_n^r}}},
\label{eqn:An_an}
\end{equation}
and imposing $[\ha_n,\had_n] = 1$, we get (after some algebraic simplifications) the commutator for the $j^\text{th}$ ion to be
\begin{equation}
[x_j, p^x_j]= i m \sum_n \frac{l_{0n}^2}{\dip{u_n^r}{u_n^r}} (2\omega_n u_{nj}^{x*}u_{nj}^x - \omega_c' \Im{u_{nj}^{y*}u_{nj}^x}).\label{eqn:x_commutator}
\end{equation}
This is further simplified by making use of the eigenvalue equations~(\ref{eqn:eig}): We define $\dKet{1^x_j}$ as $\dip{1^x_j}{u^r_n} = u^x_{nj}$, and left-multiply the second equation of Eq.~(\ref{eqn:eig}) with $\dBra{1^x_j}$. Now, using the fact that $\dmel{1^x_j}{\mathbb{L}}{u_n^r}=\omega_c' u^y_{nj}$, we get after simplification and complex conjugation,
\begin{equation}
-\dmel{1^x_j}{\K}{u_n^{r*}} - m\omega_n \omega_c' u_{nj}^{y*} = -m\omega_n^2 u^{x*}_{nj}.
\end{equation}
Multiplying both sides by $u^{x}_{nj}$ and taking the real part gives
\begin{equation}\label{eqn:K_transform}
m\omega_n \omega_c' \Im{u_{nj}^{x}u_{nj}^{y*}} = m\omega_n^2 u_{nj}^{x}u^{x*}_{nj} - \Re{u_{nj}^{x}\dmel{1^x_j}{\K}{u_n^{r*}}}.
\end{equation}
Substituting this, along with Eq.~(\ref{eqn:l0}) and $\dev{\mathbb{E}}{u^r_n} = m\omega_n^2 (1+R_n) \dip{u^r_n}{u^r_n}$, into Eq.~(\ref{eqn:x_commutator}) we can write the commutator in a more physically intuitive form,
\begin{align}
    [x_j, p^x_j] &= i\hbar \sum_n \frac{m\omega_n^2 \abs{u_{nj}^x}^2 + \Re{u_{nj}^{x}\dmel{1^x_j}{\K}{u_n^{r*}}}}{\dmel{u_n^r}{\mathbb{E}}{u_n^r}}.
    \label{eqn:x_commutator_2}
\end{align}
Intuitively, each expression in the sum can be thought to consist of the energy $E_{jn}^x$ of the $x_j,p_j^x$ degrees of freedom of ion $j$ in mode $n$ in the numerator, i.e. 
\begin{align}
E_{jn}^x = \frac{m\omega_n^2 \abs{u_{nj}^x}^2 + \Re{u_{nj}^{x}\dmel{1^x_j}{\K}{u_n^{r*}}}}{\dip{u_n^r}{u_n^r}},
\end{align}
and the total energy of the mode $E_n=\dmel{u_n^r}{\mathbb{E}}{u_n^r}/\dip{u_n^r}{u_n^r}$ in the denominator. In order for the canonical commutation relation $[x_j,p_j]=i\hbar$ to hold $\forall\; j$, we require to show that
\begin{align}
    \sum_{n} \frac{E_{jn}^x}{E_n} = 1 \; \forall\;j. 
    \label{eqn:energy_frac}
\end{align}
By energy conservation, the total energy in the $x_j,p_j^x$ degrees of freedom of ion $j$ can be written as 
\begin{align}
    E_j^x = \sum_n E_{jn}^x = \sum_n \frac{E_{jn}^x}{E_n} E_n,
    \label{eqn:energy_cons}
\end{align}
An equation like~(\ref{eqn:energy_cons}) can be proved using the equipartition theorem, where each degree of freedom---$x_j,p_j^x$ --- of an ion contributes $k_B T/2$ of energy at temperature $T$, so that the thermal average $\ev{E_j^x}=k_B T$, whereas in the language of normal modes,  every normal mode $n=1,\ldots,3N$ has $\ev{E_n}=k_B T$. In that case, taking a thermal average of Eq.~(\ref{eqn:energy_cons}) immediately proves Eq.~(\ref{eqn:energy_frac}). 

The above argument is somewhat heuristic because of our intuition-based interpretation of the energies $E_{jn}^x$ and $E_j^x$. Below, we proceed to rigorously use the equipartition theorem to prove Eq.~(\ref{eqn:energy_frac}).   

We apply the equipartition of energy on the linearized classical Hamiltonian written (i) in terms of the normal modes, Eq.~(\ref{eqn:ham_classical}), and (ii) on the linearized classical Hamiltonian written in terms of ion positions and canonical momenta,
\begin{multline}
    H = \frac{1}{2} \dev{\K}{\delta r} + \frac{1}{2m} \sum_j\left[ \left(p^x_j + \frac{m \omega_c'}{2} y_j\right)^2\right. \\\left.+ \left(p^y_j - \frac{m \omega_c'}{2} x_j\right)^2 + \left(p^z_j\right)^2\right].\label{eqn:ham_classical_linearised}
\end{multline}
Using these two forms, we find the distribution of energy in two different sets of variables, $x_j$, and $\abs{A_n}^2$ respectively.

Applying the equipartition theorem on with respect to $\abs{A_n}^2$ we find that the thermal average of the squared amplitudes are given by
\begin{equation}
\ev{\abs{A_n}^2} = \frac{k_B T}{\dev{\mathbb{E}}{u_n^r}}. \label{eqn:thermal_avg_amplitude}
\end{equation}

Now, applying the equipartition theorem on Eq.~(\ref{eqn:ham_classical_linearised}) with respect to $x_j$, we have $\ev{x_j \pdv{H}{x_j}} = k_B T$, or more explicitly,
\begin{multline}
    k_B T = \sum_n \ev{\abs{A_n}^2} \left(2 \Re{u_{nj}^{x}\dmel{1^x_j}{\K}{u_n^{r*}}} \right.\\
    \left.+ m\omega_c'\omega_n \Im{u_{nj}^x u_{nj}^{y*}}  \right).
\end{multline}
This expression on simplifying using Eq.~(\ref{eqn:K_transform}) and~(\ref{eqn:thermal_avg_amplitude}) directly leads to Eq.~(\ref{eqn:energy_frac}). This gives us the necessary canonical commutation relation,
\begin{equation}
[x_j, p^x_j] = i\hbar,
\end{equation}
and shows that the substitution, Eq.~(\ref{eqn:An_an}), is a consistent way to quantize the modes.

\section{Calculations for bilayer Ising models}\label{app:Ising}

The unitary operator describing the evolution under Eq.~(\ref{eqn:hodf_2}) is known to be analytically integrable, and forms the workhorse for implementing quantum gates with trapped ions. In this appendix, we generalize this calculation for the case of complex eigenvectors and arbitrary 3D crystals, which is relevant for the study of Ising models realized with bilayer crystals (Sec.~\ref{sec:tunable_ising}). Subsequently, we derive an approximate mean-field expression for the $\ket{\uparrow}$ state population at the end of the protocol described in Fig.~\ref{fig:tipping}(a). Finally, we derive the effective Ising coupling between ion pairs when two normal modes are simultaneously addressed using a two-tone ODF (Fig.~\ref{fig:inter_intra}).

\subsection{Derivation of effective Ising coefficients}\label{app:ising_derivation}

The unitary operator $\hUo(t)$ corresponding to the Hamiltonian~(\ref{eqn:hodf_2}) can be computed using the Magnus expansion as 
\begin{align}
 \hUo(t) &= \text{exp}\left(-\frac{i}{\hbar} \int_0^t \hHo(t_1) \dd t_1 \right.\nonumber\\
 &\left.- \frac{1}{2\hbar^2} \int_0^t \dd t_1 \int_0^{t_1} \dd t_2 [\hHo(t_1), \hHo(t_2)]\right).\label{eqn:Uodf1}
\end{align}
This expression is exact, and not approximate, because for this Hamiltonian, higher commutators of the form $[\hHo(t_1), [\hHo(t_2), \hHo(t_3)]]$ vanish. Computing the integrals in Eq.~(\ref{eqn:Uodf1}), we obtain the familiar form of $\hUo(t)$ factorized into a spin-motion unitary and a spin-spin unitary, i.e. 
\begin{align}
    \hUo(t) = \hU_\mathrm{sm}(t) \hU_\mathrm{ss}(t),
    \label{eqn:Uodf_form}
\end{align}
where 
\begin{align}
    \hU_\mathrm{sm}(t) = \exp[\sum_{j,n} (\alpha_{nj}\had_n - \alpha_{nj}^* \ha_n)\hSig^z_j],
\end{align}
and 
\begin{align}
    \hU_\mathrm{ss}(t) = \exp[-i\sum_{j\ne k} \Theta_{jk}^{zz} \hSig^z_j \hSig^z_k].
\end{align}
The spin-motion and spin-spin coupling coefficients are respectively given by
\begin{align}
	\alpha_{nj} &= \frac{F_0 l_{0n} u_{nj}^{z*}}{2\hbar\delta_n \sqrt{\dip{u^r_n}{u^r_n}}} e^{-i\phi_j} (e^{-i t \delta_n} - 1),\\
	\Theta_{jk}^{zz} &= \sum_n \frac{F_0^2 l_{0n}^2 \mathcal{U}_{n,jk}}{4\hbar^2 \delta_n^2} \left( \delta_n t \cos(\sD_{n,jk})  \right.\nonumber\\
	&\left.- \sin(\delta_n t + \sD_{n,jk}) + \sin(\sD_{n,jk}) \right)\label{eqn:Jij},
\end{align}
where $\delta_n = \mu_r - \omega_n$ is the detuning of the ODF difference frequency from the $n$th normal mode, and $\mathcal{U}_{n,jk},\sD_{n,jk}$ are defined according to Eqs.~(\ref{eqn:u_njk}) and~(\ref{eqn:d_njk}). In deriving these expressions, we have used a rotating-wave approximation to neglect rapidly oscillating terms at the frequencies $2\omega_n, 2\mu_r$ and  $\mu_r + \omega_n$. We note that these approximations break down when coupling to the low-frequency $\exb$ modes (which have a non-zero $z$ component in bilayer crystals) and hence these expressions need to be derived in full generality in that case. 

Furthermore, when the ODF is tuned close to a particular mode $n$, such that $\abs{\delta_n}\ll \abs{\delta_k}\;\forall\; k\neq n$, the effects of the other modes $k$ can be neglected in analyzing the dynamics under $\hUo(t)$. In this single-mode approximation, we can observe that the spin motion coupling vanishes ($\alpha_{nj} = 0$) at special `decoupling' times $\tau$ satisfying $\delta_n \tau = 2p\pi$ for $p\in\Z$. At these times, the effective dynamics under $\hUo(t)$ appears as an effective spin-spin interaction with coefficients  
\begin{equation}\label{eqn:theta_jk_decoup}
	\Theta_{jk}^{zz}(\tau) = \frac{F_0^2 l_{0n}^2 \mathcal{U}_{n,jk}}{4\hbar^2 \delta_n} \tau \cos(\sD_{n,jk}).
\end{equation}
In particular, if the mode mediating the interactions is predominantly real ($\mathcal{I}_n\approx 0$), such as the center-of-mass or breathing mode, then
\begin{equation}
	\Theta_{jk}^{zz}(\tau) \approx \frac{F_0^2 l_{0n}^2 \mathcal{U}_{n,jk}}{4\hbar^2 \delta_n} \tau \cos(\Delta \phi_{jk}),
\end{equation}
which demonstrates the proportionality to $\cos(\Delta\phi_{jk})$
 highlighted in Eq.~(\ref{eqn:jjk_delphi}) with $J_{jk}^{zz} = \Theta_{jk}^{zz}/\tau$.

\subsection{Analysis of protocol described in Fig.~\ref{fig:tipping}}
\label{app:tipping}

The probability to find ion $j$ in $\ket{\uparrow}$ at the end of the protocol shown in Fig.~\ref{fig:tipping} is given by $P_j = (1+\ev{\hSig_j^z})/2$, where $\ev{\hSig_j^z} = \mel{\psi_f}{\hSig_j^z}{\psi_f}$ and $\ket{\psi_f}$ is the final state at the end of the protocol, given by 
\begin{align}
    \ket{\psi_f} =  R^x_{\pi/2} \hUo(\tau) R^y_\theta\ket{\downarrow}^{\otimes N},
    \label{eqn:psif}
\end{align}
and $R^\alpha_\theta$ is a rotation about axis $\alpha=x,y,z$ with angle $\theta$. In writing Eq.~(\ref{eqn:psif}), we have assumed that $\tau$ is a decoupling time so that only the $\hU_\mathrm{ss}$ part of $\hUo$ contributes and the initial motional state can be ignored for the analysis. Analytically evaluating $\ev{\hSig_j^z}$, we find
\begin{align}\label{eqn:evSz_exact}
	&\ev{\hSig^z_j} = \mathcal{S}_j \sin(\theta),\nonumber\\
    &\mathcal{S}_j = \Im{\prod_{k, k\ne j}\left(\cos(4\Tilde{\Theta}^{zz}_{kj}) + i\sin(4 \Tilde{\Theta}^{zz}_{kj})\cos(\theta)\right)},
\end{align}
where $\Tilde{\Theta}^{zz}_{jk} = (\Theta^{zz}_{jk} + \Theta^{zz}_{kj})/2$ is the symmetric part of the coupling between ions $j,k$.  

\subsubsection{Mean-Field spin precession}

The expression in Eq.~(\ref{eqn:evSz_exact}) is exact, nevertheless, it is possible to simplify it in the mean-field limit $4\Tilde{\Theta}^{zz}_{kj} \ll 1/\sqrt{N}$ to an expression that is physically intuitive. This can be done by first multiplying and dividing each term in the product appearing in Eq.~(\ref{eqn:evSz_exact}) by the corresponding magnitude, to obtain 
\begin{multline}
    \cos(4\Tilde{\Theta}^{zz}_{kj}) + i\sin(4 \Tilde{\Theta}^{zz}_{kj})\cos(\theta)\\
    =  \exp[i\sin[-1](\tan(4 \Tilde{\Theta}^{zz}_{kj}) \cos(\theta))]\times\\
    \sqrt{\cos[2](4\Tilde{\Theta}^{zz}_{kj}) + \sin[2](4 \Tilde{\Theta}^{zz}_{kj})\cos[2](\theta)}.
\end{multline}
For $4\Tilde{\Theta}^{zz}_{kj}\ll 1$, the exponential is approximately $\exp[4i \Tilde{\Theta}^{zz}_{kj} \cos(\theta)]$, while the magnitude is approximately $\gamma_{kj} = 1 - 8 (\Tilde{\Theta}^{zz}_{kj})^2 \sin[2](\theta)$. Taking the product over ion pairs leads to 
\begin{equation}
	\mathcal{S}_j \approx \Im{\exp(\sum_{k,k\ne j} i(4\Tilde{\Theta}_{kj}\cos(\theta)))} \left(\prod_{k,k\ne j} \gamma_{kj}\right).
\end{equation}
For $4\Tilde{\Theta}^{zz}_{kj}\ll 1/\sqrt{N}$, we note that the product of $\mathcal{O}(N)$ $\gamma_{kj}$ factors is approximately $1$. As a result, we obtain 
\begin{equation}
    \mathcal{S}_j \approx \sin(\sum_{k,k\ne j} 4 \Tilde{\Theta}_{kj}\cos(\theta)),
\end{equation}
which when substituted into Eq.~(\ref{eqn:evSz_exact}) recovers the mean-field formula~(\ref{eqn:Pj_mean_field}).

\subsection{Ising coefficients under two-tone ODF}\label{app:MultimodeTuning}

It is straightforward to extend the analysis in Appendix~\ref{app:ising_derivation} to account for the effect of two tones applied to couple to different normal modes (Sec.~\ref{sec:dynamic}). Similar to the case of a single tone, the Magnus expansion for the unitary operator corresponding to Hamiltonian~(\ref{eqn:ham_two_odf}) truncates at second order, giving us the same form for the unitary as in Eq.~(\ref{eqn:Uodf_form}), but with modified coefficients given by
\begin{align}
	\alpha_{nj,\mathrm{tot}} &= \alpha_{0,nj} + e^{-i\phi_0}\alpha_{1,nj},\\
	\Theta^{zz}_{jk,\mathrm{tot}} &= \Theta^{zz}_{0,jk} + \Theta^{zz}_{1,jk} \nonumber\\&+ \sum_n \mathcal{O}\left(\frac{1}{\delta_{0,n}\delta_{1,n}} + \frac{1}{\delta_{0,n}\Delta} + \frac{1}{\delta_{1,n}\Delta}\right),
 \label{eqn:alpha_jij_twotone}
\end{align}
where $\alpha_{\gamma,nj},\Theta_{\gamma,jk}^{zz}$ ($\gamma=0,1$) are the coefficients obtained when only the tone $\gamma$ is applied, and $\phi_0$ is a relative phase between the two tones. Furthermore, $\Delta=\mu_{r,0} - \mu_{r,1}$ is the difference between the two ODF difference frequencies corresponding to the two tones, and $\delta_{\gamma,n} =\mu_{r,\gamma} - \omega_n$ is the detuning of the tone $\gamma$ from the $n$th normal mode. In Sec.~(\ref{sec:dynamic}), we tune the $\gamma=0,1$ tones close to the c.m. and the breathing modes respectively, which we label as $n=0$ and $n=1$ for the present discussion. In this case, $\abs{\delta_{0,0}},\abs{\delta_{1,1}}\ll \abs{\Delta},\abs{\delta_{0,1}},\abs{\delta_{1,0}}$, because we typically choose $\abs{\delta_{0,0}}=\abs{\delta_{1,1}}\sim 2\pi\times 1$ kHz, whereas the c.m. and breathing modes for the crystal used in Fig.~\ref{fig:inter_intra} are separated by approximately $2\pi\times 154$ kHz. Hence, the error due to the cross terms shown as small corrections in Eq.~(\ref{eqn:alpha_jij_twotone}) are expected to be $100\times$ smaller than the effective magnitude of $\Theta_{jk,\mathrm{tot}}^{zz}$, and hence it can be well approximated as the sum of the individual spin-spin coupling coefficients coming from the coupling of tone $0$ to the c.m. mode and tone $1$ to the breathing mode. Finally, comparing Eq.~(\ref{eqn:alpha_jij_twotone}) to Eq.~(\ref{eqn:theta_jk_decoup}) and Eq.~(\ref{eqn:Jij_tunable}), we obtain that
\begin{align}
    C_{jk} &= \frac{l_{00}^2 \mathcal{U}_{0,jk}}{4\hbar^2} \cos(\sD_{0,jk}),\nonumber\\
    D_{jk} &= \frac{l_{01}^2 \mathcal{U}_{1,jk}}{4\hbar^2} \cos(\sD_{1,jk}),
    \label{eqn:cjk_djk}
\end{align}
since $\delta_{0,0} = \delta_0$ and $\delta_{1,1} = \delta_1$.

In the case of $\alpha_{nj,\mathrm{tot}}$, similar arguments using the relative magnitudes of various detunings show that $\abs{\alpha_{0,1j}},\abs{\alpha_{1,0j}}\ll 1$ and hence can be neglected at all times. Furthermore, since we set $\abs{\delta_{0,0}}=\abs{\delta_{1,1}}$, $\alpha_{0,0j},\alpha_{1,1j}$ vanish at identical decoupling times, given by $\tau=2p\pi/\abs{\delta_{0,0}}$ for $p\in \Z$.

\section{Derivation of spin exchange model}\label{app:ChiralModels}
In this appendix, we derive the effective spin exchange model obtained when the ODF interaction is applied in the presence of an additional transverse field. In an interaction picture with respect to the free Hamiltonian of the spins and the normal modes, the interaction Hamiltonian for the system in the presence of a transverse field that is resonant with the $\ket{\downarrow}\to\ket{\uparrow}$ transition is given by Eq.~(\ref{eqn:ham_transverse_odf}), which we repeat below for convenience:
\begin{multline}
\hat{H}_\mathrm{int.} =  \sum_j \frac{\hbar B_0}{2} \hSig^x_j + \sum_j F_0 \cos(\mu_r t + \phi_j) \hat{z}_j(t) \hSig^z_j.
\label{eqn:ham_transverse_odf_app}
\end{multline}
To derive an effective spin model, we expand $\hat{z}_j(t)$ according to Eq.~(\ref{eqn:delta_r_quantized}) and transform into a second interaction picture with respect to the Hamiltonian  $\sum_j (\hbar B_0/2) \hSig^x_j$. Furthermore, we work in a rotated spin space for each spin by introducing operators $\hat{\tau}_j^x\equiv \hSig_j^z$ and $\hat{\tau}_j^z\equiv -\hSig_j^x$, to obtain the Hamiltonian 
\begin{multline}
	\hat{H}_\mathrm{int.}^{(2)} = 
    \sum_{j,n} \frac{F_0 l_{0n}}{\sqrt{\dip{u^r_n}{u^r_n}}} \cos(\mu_r t + \phi_j)\times\\(\hat{a}_n e^{-i\omega_n t} u_{nj}^z + \hat{a}_n^\dagger e^{+i\omega_n t} u_{nj}^{z*})\times\\(\hat{\tau}^+_j e^{-iB_0 t} + \hat{\tau}^-_j e^{iB_0 t}).
    \label{eqn:ham_transverse_odf_app_2}
\end{multline}
We write the $\cos$ term as a complex exponential and expand the product in Eq.~(\ref{eqn:ham_transverse_odf_app_2}). Next, for $\abs{B_0}\ll  \mu_r,\omega_n$, which is the case in our work, we can neglect terms rotating at $\mu_r + \omega_n \pm B_0$ since they are rapidly oscillating. We then obtain the Hamiltonian
\begin{multline}\label{eqn:ham_transverse_odf_app_3}
	\hat{H}_{I} \approx 
    \sum_{j,n} \frac{F_0 l_{0n}}{2\sqrt{\dip{u^r_n}{u^r_n}}} \left(
\hat{a}_n^\dag \hat{\tau}^+_j u_{nj}^{z*} e^{-i[(\delta_n+B_0)t+\phi_j]}\right.\\ \left.
+\hat{a}_n^\dag \hat{\tau}^-_j u_{nj}^{z*} e^{-i[(\delta_n-B_0)t+\phi_j]} +\text{h.c.} \right),
\end{multline}
where $\delta_n = \mu_r - \omega_n$ as before. This Hamiltonian forms the starting point to derive an effective spin model using effective Hamiltonian theory~\cite{james2007CJP}. In this approach, starting from a Hamiltonian with harmonic terms of the form 
\begin{equation}
	\hH = \sum_n \hh_n e^{-i\nu_n t} + \hhd_n e^{i\nu_n t},
\end{equation}
and coarse-graining the corresponding unitary operator over fast timescales $\nu_n^{-1},(\nu_n+\nu_k)^{-1}$, we can obtain an effective Hamiltonian of the form 
\begin{equation}
	\hH_\text{eff} = \sum_{n,k} \frac{\nu_n + \nu_k}{2\hbar \nu_n \nu_k} [\hhd_k, \hh_n] e^{i(\nu_k - \nu_n)t}.
\end{equation}
Applying this formula to the Hamiltonian~(\ref{eqn:ham_transverse_odf_app_3}) and assuming that the normal modes are initially in vacuum, so that $\had_m \ha_n \rightarrow 0\;\forall\;m,n$, we obtain 
\begin{align}
    \hH_\text{eff} = &\sum_{j} \frac{H_j}{2} \hTau_j^z + \sum_{j \neq k} J_{jk}^\text{ff} \hTau_k^+ \hTau_j^- \nonumber\\
    &\sum_{j \neq k} \left(J_{jk}^\text{pp} \hTau_k^+ \hTau_j^+ e^{-2iB_0 t} + \text{h.c.}\right),    
\end{align}
where the self-energy and flip-flop coefficients $H_j,J_{jk}^\text{ff}$ are given in Eqs.~(\ref{eqn:hj}),(\ref{eqn:jkj_ff_1}) and (\ref{eqn:jkj_ff_2}). This effective spin model is valid provided $NJ^\text{ff},NJ^\text{pp}\ll \abs{\delta_n\pm B_0}$, where $J^\text{ff},J^\text{pp}$ are typical values of the respective coefficients. The $J_{jk}^\text{pp}$ terms correspond to a pair-production (pp) and annihilation of spin excitations. They are given by 
\begin{equation}
    J_{jk}^\text{pp} = \sum_{n} \frac{F_0^2 l_{0n}^2\delta_n\mathcal{U}_{n,kj}}{4\hbar^2 (\delta_n^2-B_0^2)}\cos(\mathcal{D}_{n,kj}).
\end{equation}
However, these terms are oscillating at $\pm 2B_0$. Therefore, if additionally $NJ^\text{pp}\ll 2\abs{B_0}$, these terms are also rapidly oscillating and can be neglected in a first approximation.

When the modes are at finite temperatures with mean occupations $\nbar_n$, the normal mode operators cannot be removed from the self-energy terms in general. Specifically, $H_j$ becomes time-dependent with contributions from terms of the form $\had_m \ha_n e^{i(\delta_n - \delta_m)t}$. However, if the modes in the motional spectrum are well-separated, we can retain only the $m=n$ terms, and replace $\had_n \ha_n\to\nbar_n$, to obtain 
\begin{equation}
	H_j = \sum_n \frac{F_0^2l_{0n}^2\mathcal{U}_{n,jj}B_0 (2\nbar_n + 1)}{2\hbar^2 (\delta_n^2-B_0^2)},
\end{equation}
which is once again independent of the mode operators. On the other hand, the spin exchange coefficients $J_{jk}^\text{ff}$ are independent of temperature and remain unchanged.

\end{document}